# EIGENRAYS IN 3D HETEROGENEOUS ANISOTROPIC MEDIA: PART III – KINEMATICS, FINITE-ELEMENT IMPLEMENTATION

*Zvi Koren and Igor Ravve (corresponding author), Emerson*

zvi.koren@emerson.com   ,    igor.ravve@emerson.com

## ABSTRACT

Following the theory presented in Part I, where we derived the Euler-Lagrange, nonlinear, second-order kinematic (boundary-value, two-point) ray tracing equation for smooth heterogeneous general anisotropic media, this part is devoted for its numerical finite-element solution. For a given initial-guess (non-stationary) trajectory between two fixed endpoints, discretized with a set of (non-uniformly spaced) nodes, we update the location and direction of the ray trajectories at the nodes (the degrees of freedom of the finite element method), to obtain the nearest stationary ray path.

Starting with the Euler-Lagrange equation derived in Part I, we apply the weak formulation and the Galerkin method to reduce this second-order, ordinary differential equation into a nonlinear, local, first-order, weighted residual algebraic equation set. The solution is based on a finite element approach with the Hermite polynomial interpolation, for computing the ray coordinates, directions and the traveltime between the nodes. The Hermite interpolation is a natural choice, since, in addition to the nodal locations, also the ray directions at the nodes (the arclength derivative of the nodal locations) are independent degrees of freedom of the finite element method. This is in particular important in anisotropic media where the ray directions coincide with the ray velocity directions that constitute the Lagrangian (time integrand) along the ray.

We then construct the target function to be minimized, where we distinguish between two functions triggered by the two possible types of stationary rays: minimum traveltime and saddle-point solutions, mainly in cases where the stationary paths include caustics. The target functions include two penalty terms related to two essential constraints. The first is related to the distribution of the nodes along the ray and the second enforces the normalization of the ray direction to a unit length at each node along the ray. The minimization process involves the computation of the traveltime gradient vector and the Hessian matrix. The latter is used for the minimization process and to analyze the type of the stationary ray. Finally, we demonstrate the efficiency and accuracy of the proposed method along three canonical examples.

## INTRODUCTION

A comprehensive review of two-point boundary-value ray tracing methods in isotropic and anisotropic elastic media, and in particular of the ray bending method (which is the class of solutions that includes the proposed Eigenray method), has been provided in Part I of this study. This part is a direct continuation of Part I, where we derived the Euler-Lagrange, nonlinear, second-order kinematic ordinary differential equation for the stationary ray path between two fixed endpoints. We proposed a specific Lagrangian which depends on the ray location coordinates and the ray directions, where both locations and directions are functions of the arclength used as the flow variable. In Part II, we validated the correctness of the proposed Lagrangian and showed its relation and consistency with alternative Lagrangians for heterogeneous general anisotropic elastic media.

In this part, given an initial-guess (non-stationary) trajectory, we find the stationary ray path using the finite element approach. The initial trajectory is discretized by a number of segments (finite elements) whose lengths are normally shorter in regions of high velocity gradients. The finite elements consist of either two or three nodes, depending on the required accuracy. In each elementary interval $ds$ along the path, the traveltime $d\tau$ (and hence also the Lagrangian, $L = d\tau/ds$) depend on the ray (group) velocity magnitude, which in anisotropic media, is a function of both, the location and the direction of propagation. It is therefore natural to apply the Hermite interpolation, which is based on the locations of the nodes, $\mathbf{x}_i(s)$, and the nodal directions, $\mathbf{r}_i(s) = \dot{\mathbf{x}}_i(s) = d\mathbf{x}(s)/ds\big|_i$, of the path. The arclength, $s$, is the flow parameter, and the nodal ray directions are the arclength derivatives of the locations. The Hermite interpolation is normally used to guarantee the continuity of the path points' location and the continuity of the ray direction. The latter, however, can be controlled and managed to enforce discontinuities whenever necessary, for example, in "blocky" layer-model representations, at the interfaces between two given layers with different elastic properties.

In this study we present two different approaches that yield the same algebraic equation set, to be solved numerically (iteratively) using the Newton method (or a gradient-based method), for converging into the stationary ray path. The first approach involves direct search for the vanishing of the spatial and directional components of the traveltime gradient vector. The second approach is based on the application of the weak formulation and the Galerkin (1915) method to the Euler-Lagrange, nonlinear, second-order kinematic equation (equation 8 in Part I). It leads to a nonlinear, local, first-order, weighted residual algebraic equation set. "First-order" means that the weak formulation eliminates the second derivatives, $\ddot{\mathbf{x}}(s) = \dot{\mathbf{r}}(s)$, of the location coordinates with respect to (wrt) the arclength $s$ (i.e., the curvature vector), and only the locations, $\mathbf{x}(s)$, and their first derivatives, $\dot{\mathbf{x}}(s) = \mathbf{r}(s)$, (i.e., the ray directions) remain. By using the same interpolation method (in this study, the Hermite polynomial interpolation), the local algebraic equation sets obtained from the two approaches are equivalent. Finally, assembling the local equations together in a unique set and applying the boundary conditions at the endpoints of the path, we obtain the resolving nonlinear algebraic system for the whole ray trajectory.

The stationary path may deliver either a minimum traveltime or a saddle-point solution in case the path contains caustics. The solution of the resolving equation set is obtained using an iterative procedure. In the case of a minimum traveltime, we apply the Newton optimization method, minimizing the target function that consists of the traveltime augmented by two essential constraints. In the case of a saddle point or in a general case when the type of the stationary path is not known ahead, a gradient-based method is used, where the target function to be minimized consists of the traveltime gradient squared augmented by the abovementioned constraints. We note that in either case, the target function is to be minimized, and both the global traveltime gradient vector and the global traveltime Hessian matrix should be computed. The local gradients and Hessians of the traveltime, in turn, depend on the corresponding derivatives of the ray velocity magnitude, which are the actual core computations of this work. A detailed description of these derivatives is given in Ravve and Koren (2019) and a brief summary in Appendix E of Part I.

The target function to be minimized includes penalty terms related to two different constraints. The first constraint is related to the distribution of the nodes along the ray, enforcing denser distribution at high local curvatures of the ray path. The second constraint enforces the normalization of the ray direction to a unit length, $\mathbf{r} \cdot \mathbf{r} = 1$, at each node along the ray.

As mentioned, with the proposed Hermite-based finite element method, one can, for example, enforce harsh discontinuities of both the ray velocity magnitude and its direction at the interfaces. The discontinuous ray velocity direction does not affect the local traveltime gradient vectors and Hessian matrices of the finite elements. With the Hermite interpolation, where the ray directions are separate DoF, these discontinuities are implemented at a later stage, when we assemble the local traveltime gradients and Hessians into the global ones. The ability to control the continuity of the ray direction (keeping it normally continuous, but discontinuous where necessary) is a great advantage. We consider the additional DoF provided by the Hermite elements as one of the main strengths of the proposed method.

The introduction section of Part I includes a comprehensive list of references describing different seismic ray tracing methods and in particular the ray bending approach where the Eigenray method can be considered one of its versions. Hence, in this part we only add the references related to the actual finite element implementation. Galerkin (1915) suggested the weighted residual method to approximate solutions of ordinary and partial differential equations. The method, along with other key finite-element procedures (choice of the interpolation functions, numerical integration, assembly, equation solvers, etc.) is described in many textbooks, like Segerlind (1984), Hughes (2000), Reddy (2004), Zienkiewicz et al. (2013), Bathe (2014), and others.

In this part, we demonstrate the capabilities and accuracy of the method to obtain stationary rays, considering kinematic solutions for three benchmark isotropic examples. Additionally, in Part

VII we demonstrate both kinematic and dynamic solutions for eight more examples, focusing on different anisotropic cases.

Appendices

Appendices A and B provide known background material which we find essential for the understating of this study and for the actual implementation. The other appendices, listed below, present the original theoretical materials used/derived in this study:

In Appendix A, we list the Hermite interpolation formulae for two-node and three-node finite elements.

In Appendix B, we present the numerical quadrature (integration) formulae, based on the Hermite interpolation, which are needed to compute the local traveltime, its gradient and Hessian.

In Appendix C, we derive the relationship (metric) between the arclength, $s$, and the internal flow parameter, $-1 \leq \xi + 1$, of the finite element, used for parameterizing the ray path.

In Appendix D, we refer to the so-called normalized Lagrangian, $\hat{L}(\xi)$, where the arclength flow parameter is replaced by an internal unitless flow parameter, $\xi$. This Lagrangian is used in the finite element solver. In particular, we relate the spatial/directional gradients and spatial/directional/mixed Hessians of the normalized Lagrangian to their corresponding gradients and Hessians of the arclength-related Lagrangian.

In Appendix E, we derive the relationships for the interpolated local traveltime.

In Appendix F, we obtain the first and second derivatives of the local traveltime, leading to the solution of Fermat's principle of stationary traveltime. The second derivatives are primary needed because we use the Newton optimization method. They are also essential in cases where the stationary rays involve caustics, because in these cases the gradient itself of the target function includes both the gradient and Hessian of the traveltime.

In Appendix G, we describe the finite element solver for the second-order, nonlinear, Euler-Lagrange ODE obtained in Part I. We apply the weak formulation and the Galerkin method to this ODE, and we reduce it to the first-order, local, nonlinear weighted residual equation set. We also show that this algebraic equation set is fully equivalent to the equation set obtained by the direct derivation of Fermat's principle of stationary traveltime, described in Appendix F. In other words, the weak finite-element formulation leads to the same spatial and directional sub-blocks of the local traveltime gradient as the sub-blocks obtained by the direct search for the stationary traveltime, and thus to the same local and global traveltime gradients.

In Appendix H, we present the formulae needed to compute the penalty terms of the constraints, included within the target function, and their derivatives.

In Appendix I, we explain the assembly of the local traveltime gradients and Hessians of the finite elements, and the local gradients and Hessians of the constraints, into the global (all-node) gradient vector and Hessian matrix of the entire Eigenray finite-element scheme, and implement the required boundary conditions for the end nodes. We further discuss the possibility of allowing/enforcing discontinuous ray velocity directions at specific nodes located at sharp model discontinuities (e.g., along interfaces). Continuity/discontinuity of the ray directions does not affect the local traveltime gradients and Hessians of the finite elements; this feature is related to additional direction DoF and implemented at the assembly.

In Appendix J, we show the optimization procedure that involves the Newton method and the anti-gradient descent: The latter is used for a minimum search when the former fails to reduce the value of the target function, or in the case of a saddle-point type stationary path.

In this study, we apply tensor notations and do not distinguish between row and column vectors.

## FINITE-ELEMENT DISCRETIZATION

A schematic ray path discretization is shown in Figure 1, where three-node elements are used. The fixed endpoints of the ray path are shown in green, the finite-element joints in black, and the internal nodes of the elements in red. (Obviously, a realistic scheme includes many more elements and nodes.) We define a segment of the path as an interval between two neighbor nodes. Two-node elements consist of a single segment, while three-node elements include two segments. The continuous ray trajectory $\mathbf{x}(s)$, where $s$ is the arclength flow parameter (with an infinite number of DoF) is represented (approximated) by a finite number of segments (intervals of finite elements) with nodes at their joints, $\mathbf{x}_i$. At the joint nodes connecting the neighbor elements, the locations, $\mathbf{x}(s)$, and ray directions $\mathbf{r}(s) \equiv \dot{\mathbf{x}}(s) \equiv d\mathbf{x}/ds$, related to the adjacent elements are normally continuous (except the directional discontinuities at the medium interfaces, described above). At the internal nodes of the elements, the locations, ray directions and all higher derivatives of the locations wrt the arclength are continuous. The number of DoF becomes finite: three location components $\mathbf{x}_i$ and three ray direction components $\mathbf{r}_i$, per node $i$. The three nodal direction DoF are dependent since a direction in 3D space is defined by two angles; we keep the ray direction normalized to the unit length, $\mathbf{r}_i \cdot \mathbf{r}_i = 1,$ and we enforce this dependency with the soft normalization constraint.

In this study, we apply both two-node and three-node finite elements, where the latter have an additional internal node. The resolving nonlinear algebraic equation set of the finite element analysis delivers a solution to the kinematic problem (locations and ray directions) at all nodes, and the Hermite interpolation makes it possible to compute the solution continuously for any point between the nodes.

Remark: As the arclength between each pair of successive nodes decreases, the finite-element solution converges to the exact "theoretical" solution. For the exact solution, any segment of the stationary path is also a stationary path. The same holds for the finite-element solution, at least at the limit when the arclengths of all segments become small. However, we still need to keep in mind that the (stationary) ray path within each specific finite element it represents a result of the Hermite polynomial interpolation, rather than a true solution of the kinematic ray tracing equation in heterogeneous anisotropic media.

We finally note that with the Hermite interpolation, even if the ray directions at the finite element joints are continuous, the curvatures at the element joints are discontinuous. This is important since the nodal curvature appears in the equations of the dynamic finite-element formulation. Hence, at the joints, we average the curvature values of the preceding and successive finite elements at that node.

## THE ALGEBRAIC EQUATION SET

The actual ray path delivers a stationary traveltime $t$,

$$t = \int_S^R L(\mathbf{x}, \mathbf{r}) ds = \int_S^R \frac{\sqrt{\mathbf{r} \cdot \mathbf{r}}}{v_{\text{ray}}(\mathbf{x}, \mathbf{r})} ds \to \text{stationary} \quad , \tag{1}$$

where the Lagrangian $L$ is a function of the arclength-dependent ray location, $\mathbf{x}(s)$, and ray direction, $\mathbf{r}(s) = d\mathbf{x}/ds$, and $v_{\text{ray}}$ is the magnitude of the ray velocity. We start with an initial guess trajectory between the source $S$ and receiver $R$, discretized with a number of (non-uniform) segments (intervals of finite elements). The density of the nodes along the trajectory is directly related to the local curvature of the trajectory (see Appendix H).

Local traveltime and its derivatives

The total traveltime, that has to be stationary for the resolved path, is the sum of the local traveltimes, $t = \sum \Delta t_i$. The term "local" means a characteristic within a single finite element. We use two-node or three-node finite elements, and we apply the Hermite interpolation between the nodes (Appendix A). The quadrature (integration) formulae based on this interpolation are needed to compute the traveltime and its derivatives at each element (Appendix B).

To keep the limits of integration fixed and identical for all finite elements, inside the finite elements, we introduce an internal unitless flow parameter, $\xi$, $-1 \leq \xi \leq 1$, to be used instead of the arclength $s$ in the finite-element solver. Given the values of the nodal locations $\mathbf{x}$ and directions $\mathbf{r}_i = \dot{\mathbf{x}}_i = d\mathbf{x}/ds|_i$, we minimize the pseudo strain energy within a single element in order to convert the nodal directions $\mathbf{r}_i$ into the nodal derivatives of the position wrt the internal parameter, $\mathbf{x}'_i = d\mathbf{x}/d\xi|_i$. For this we compute the nodal values of the metric, $s'_i = ds/d\xi|_i$ (see

Appendix C). Recall that in our notation, a dot symbol over a variable means a derivative wrt the arclength $s$, and a prime means a derivative wrt the internal parameter $\xi$ of a finite element.

The ray direction wrt to the internal unitless flow parameter at any point reads,

$$\mathbf{r}(\xi) = \dot{\mathbf{x}} = \frac{d\mathbf{x}}{ds} = \frac{d\mathbf{x}}{d\xi}\frac{d\xi}{ds} = \frac{\mathbf{x}'(\xi)}{s'(\xi)} = \frac{\mathbf{x}'(\xi)}{\sqrt{\mathbf{x}'(\xi)\cdot\mathbf{x}'(\xi)}} \quad . \tag{2}$$

In Appendix D, we introduce the normalized Lagrangian, where the arclength flow parameter is replaced by the internal parameter, $\xi$. In particular, we relate the spatial/directional gradients and spatial/directional/mixed Hessians of the normalized Lagrangian $\hat{L}(\xi)$ to their corresponding gradients and Hessians of the arclength-related Lagrangian $L(s)$.

Hence, for each element we compute the local traveltime (Appendix E),

$$\Delta t = \int_{\xi=-1}^{\xi=+1} \frac{\sqrt{\mathbf{x}'(\xi)\cdot\mathbf{x}'(\xi)}}{v_{\text{ray}}\left[\mathbf{x}(\xi),\mathbf{x}'(\xi)\right]} d\xi = \int_{\xi=-1}^{\xi=+1} \hat{L}(\xi) d\xi \quad , \tag{3}$$

where $\hat{L}(\xi)$ is the normalized Lagrangian (see Table 2 of Part I), i.e., the traveltime integrand wrt the unitless internal parameter $\xi$. Note that $\hat{L}$ has units of time, [T], while the arclength-related Lagrangian, $L(\mathbf{x},\mathbf{r}) = L(s)$, has units of slowness, [T/L]. The two Lagrangians are related with the metric, $ds/d\xi$,

$$\hat{L}(\mathbf{x},\mathbf{r}) d\xi = L(\mathbf{x},\mathbf{r}) ds = d\tau \quad \rightarrow \quad \hat{L}(\mathbf{x},\mathbf{r}) = L(\mathbf{x},\mathbf{r})\frac{ds}{d\xi} \quad . \tag{4}$$

Next, we compute the local traveltime derivatives wrt the nodal positions and directions: the local gradient (vector of length $6n$) and the Hessian (matrix of dimensions $6n \times 6n$), where $n$ is the number of nodes in a single element (i.e., either 2 or 3 in this study). The details for computing the local traveltime and its derivatives are explained in Appendices E and F, respectively.

Two approaches for obtaining the algebraic equation set

The finite-element formulation yields a set of nonlinear algebraic equations to be iteratively solved with the Newton or gradient methods, in order to converge to the nearest stationary ray path. Starting our derivation from equation 1, we demonstrate two different approaches for obtaining this nonlinear algebraic set.

In the first approach we directly apply the stationarity condition to the traveltime integral of equation 1, leading to the vanishing global traveltime gradient wrt locations $\mathbf{x}_i$ of the nodes and the ray direction vectors $\mathbf{r}_i$ at the nodes, where subscript $i = 0, ..., N$ is the node index, and the total number of nodes is $N+1$. The ray discretization and the Hermite interpolation make it possible to obtain the local traveltime gradients related to the individual finite elements (Appendix F), to be assembled into the global traveltime gradient vector of the whole path (Appendix I). The vanishing global traveltime gradient, along with the boundary conditions, constitutes the nonlinear algebraic kinematic equation set.

In the second approach, we follow the result obtained in Part I, where the stationarity condition is applied to the traveltime functional in equation 1 leads to the Euler-Lagrange, second-order, nonlinear ODE (equation 8 of Part I),

$$\frac{d}{ds}\left(\frac{\mathbf{r}}{v_{ray}} - \frac{\nabla_{\mathbf{r}} v_{ray}}{v_{ray}^2}\right) = -\frac{\nabla_{\mathbf{x}} v_{ray}}{v_{ray}^2} \quad . \tag{5}$$

The detailed derivation for the construction of the nonlinear algebraic kinematic equation set is presented in Appendix G. We apply the weak formulation with the Galerkin method to this ODE (equation 2), where both the interpolation and the test (weight) functions are the same Hermite polynomials. The residual of the ODE is orthogonal to each test function. "Orthogonal" means that the integral of their product within the finite-element arclength vanishes. This procedure includes integration by parts and effectively reduces the second-order nonlinear ODE set to the first-order, nonlinear, local weighted residual set. The resulting local weighted residual equations are identical to the local traveltime gradient equations (the nonlinear algebraic set) obtained with the first approach described in Appendix F. In both cases, the normalized internal variable $\xi$ (defined within each individual finite element and related to the arclength $s$ by means of the metric function) is used as the flow (characteristic) parameter.

## CONSTRAINT ON THE NODE DISTRIBUTION ALONG A STATIONARY PATH

While the stationary traveltime condition fully defines the ray path, it still allows some freedom for setting the distribution of the nodes along the ray. We therefore apply an additional constraint $W_s$ on the segment lengths between the successive nodes, so that the nodes are located more densely along ray parts with high curvature. In these parts of the path, the ray velocity magnitude changes rapidly in the direction normal to the ray velocity vector. We construct constraints on the ratios between the arclengths connecting successive nodes. These lengths are inversely proportional to the average curvatures of the corresponding intervals,

$$W_s = \frac{w_s}{2} \sum_{i=1}^{N-1} \left( \frac{\Delta s_i}{\Lambda_i} - \frac{\Delta s_{i+1}}{\Lambda_{i+1}} \right)^2 \quad , \tag{6}$$

where $w_s$ is a weighting factor of the constraint, $N$ is the number of intervals (ray path segments), $N+1$ is the total number of nodes enumerated from zero to $N$, and $N-1$ is the number of internal nodes of the path. $\Delta s_i$ is the arclength of segment $i$, and parameter $\Lambda_i$ is the mean radius of curvature of the ray along this segment, normalized so that $\sum_{i=1}^{N} \Lambda_i = 1$; $\Lambda_i$ is limited to a finite (large) value for straight segments.

In general, there are two ways to implement the constraints: hard constraints (e.g., applying the Lagrange multipliers method) and soft or relaxed constraints, by adding a penalty term to the target function to be minimized. The soft constraint method (used in this study) is simpler; it does not lead to additional unknown parameters and does not increase the bandwidth of the resolving matrix, while still providing excellent accuracy. There is no need to keep the desired length ratio exact. The details of the length constraint implementation are given in Appendix H.

## CONSTRAINT ON THE RAY DIRECTION NORMALIZATION

Throughout the workflow, the ray velocity direction in the governing equations is assumed normalized, where at each node, $\mathbf{r} \cdot \mathbf{r} = 1$. However, the Newton iterative procedure (or any other iterative procedure, like the conjugate-gradient or anti-gradient descent, that does not explicitly require the second derivatives of the target function) provides a "recommended" set of updated parameters at the end of each successive iteration, which can violate the normalization of the ray direction. For example, the direction at node $i$ becomes $\mathbf{r}_i + \Delta \mathbf{r}_i$, where $|\mathbf{r}_i| = 1$, but the length $|\mathbf{r}_i + \Delta \mathbf{r}_i|$ may essentially differ from 1. The remedy is to include within the target function an additional normalization penalty term $W_r$,

$$W_r = \frac{w_r}{2} \sum_{i=0}^{N} (\mathbf{r}_i \cdot \mathbf{r}_i - 1)^2 \quad , \tag{7}$$

where $w_r$ is a weighting factor of the constraint, and $N+1$ is the total number of nodes, including the endpoints of the path. The implementation details of this constraint are given in Appendix H.

Remark 1: The direction normalization constraint is important for the convergence of the solution to the minimum value of the target function. (Recall that this function is minimized for both types of a stationary traveltime: a minimum and a saddle point.) Without that constraint term, if we only normalize the length of the updated ray directions $\mathbf{r}_i + \Delta \mathbf{r}_i$ after each iterative step (which in any case we do), there will be essential direction differences between the normalized updates $\Delta \tilde{\mathbf{r}}_i = (\mathbf{r}_i + \Delta \mathbf{r}_i)/|\mathbf{r}_i + \Delta \mathbf{r}_i| - \mathbf{r}_i$ and the non-normalized corrections $\Delta \mathbf{r}_i$, obtained by the Newton iteration. Note that both, the old and the updated, nodal ray directions are normalized, $|\mathbf{r}_i| = 1$ and $|\mathbf{r}_i + \Delta \tilde{\mathbf{r}}_i| = 1$. However, this operation alone does not suffice. The discrepancy, $\Delta \tilde{\mathbf{r}}_i - \Delta \mathbf{r}_i$, significantly decreases the efficiency of the Newton method. Our computational practice confirms that by introducing the normalization penalty term as part of the target function, these differences become small for all nodes. We further note that these small differences are acceptable and always exist because the applied normalization constraint is soft.

Remark 2: We emphasize that this normalization penalty should not be confused with the normalized directional gradient of the ray velocity $\nabla_\mathbf{r} v_{\text{ray}}$, and the directional and mixed Hessians of the ray velocity, $\nabla_\mathbf{r} \nabla_\mathbf{r} v_{\text{ray}}, \nabla_\mathbf{x} \nabla_\mathbf{r} v_{\text{ray}}, \nabla_\mathbf{r} \nabla_\mathbf{x} v_{\text{ray}}$, whose components are also normalized. Normalization of the partial derivatives of the ray velocity means that in order to keep the unit length of the ray direction vector $\mathbf{r}$, one cannot independently vary one of the ray direction components without simultaneously applying the corresponding compensational changes to the two other components (even when this variation is infinitesimal).

## ASSEMBLY OF THE LOCAL TRAVELTIME GRADIENTS AND HESSIANS

The target function to be optimized includes the traveltime and two soft weighted constraints related to the distribution of nodes along the ray path and to the normalization of the directions of the ray velocity at the nodal points. Each element contributes to the traveltime, each joint node contributes to the distribution penalty, and each node contributes to the normalization penalty. The derivative of the sum equals to the sum of the derivatives, and this means that at the joints, where the local vectors and matrices overlap, the numbers are just added. Locations of the end nodes are known, and this constitutes the boundary conditions. The details of the assembly approach and the implementation of the boundary conditions at the end nodes are explained in Appendix I. In this appendix, we also discuss the possibility of allowing (or enforcing) discontinuous ray directions through areas of sharp velocity variations (e.g., sharp transition zones or specified interfaces).

## OPTIMIZATION OF THE TARGET FUNCTION

As mentioned, in the case of a minimum traveltime, the target function $T$ includes the traveltime and two weighted penalty terms. The target function and its gradient read,

$$T = t + W_s + W_r \to \min \quad , \quad \nabla_{\mathbf{d}} T = \nabla_{\mathbf{d}} t + \nabla_{\mathbf{d}} W_s + \nabla_{\mathbf{d}} W_r = 0 \quad , \quad (8)$$

where $t$ is the traveltime, symbol $\nabla_{\mathbf{d}}$ means the gradient of a scalar function wrt all (location and direction) DoF, $W_s$ is the node distribution penalty, and $W_r$ is the ray direction normalization penalty. Theoretically, the stationary path may also deliver a maximum traveltime; however, in real problems this case is very rare (non-realistic) – it requires all eigenvalues of the traveltime Hessian matrix to be negative. On the other hand, the case of stationary traveltime with saddle points is very common, indicating caustic locations along the ray. In this case or in cases when the type of stationary point is unknown, we suggest that the target function consists of the traveltime gradient squared (instead of the traveltime) and the penalty terms, where the minimum of the target function corresponds to the vanishing (or negligibly small) norm of the gradient vector. For a general or saddle-point case, the target function and its gradient read,

$$T = \frac{\nabla_{\mathbf{d}} t \cdot \nabla_{\mathbf{d}} t}{2} + W_s + W_r \to \min \quad , \quad \nabla_{\mathbf{d}} T = \nabla_{\mathbf{d}} \nabla_{\mathbf{d}} t \cdot \nabla_{\mathbf{d}} t + \nabla_{\mathbf{d}} W_s + \nabla_{\mathbf{d}} W_r \quad , \quad (9)$$

where $\nabla_{\mathbf{d}} t$ and $\nabla_{\mathbf{d}} \nabla_{\mathbf{d}} t$ are respectively the gradient and Hessian of the traveltime wrt all DoF. We take into account that the location DoF have the units of distance, while the direction DoF are unitless. Note that for this general case of the stationary traveltime, in particular, for a saddle-point case when equation 9 is applied, the target function itself includes the traveltime gradient, and as a result, the gradient of the target function already includes the traveltime Hessian. Therefore, in these cases, we do not apply the Newton-type minimization methods that require the second derivative of the target function (and hence, higher order derivatives of the traveltime), and we only apply the gradient methods (e.g., the conjugate-gradient or anti-gradient descent). We further note that a) in either case, whether the traveltime of the stationary path is the minimum or not, the target function is always minimized, and b) for either of the minimization methods that we use (Newton-type or gradient), we need to compute both the gradient and the Hessian of the traveltime.

Comment: Convergence to a zero traveltime gradient squared (to the stationary ray path) may be sensitive to the initial-guess trajectory. If the initial guess is far from the stationary solution, the iterations may yield a non-vanishing minimum traveltime gradient squared.

For the traveltime minimum search, we use the Newton optimization method, where at each iteration of the optimization process, we update the nodal locations and directions of the ray and refine the trial trajectory until the stationary condition is reached. Upon completion of the iterative procedure, the penalty terms accept marginally small values, so that the stationary condition of the entire target function is fairly close to the traveltime stationarity. The details are presented in Appendix J.

## NUMERICAL EXAMPLES

We present a number of synthetic examples for the proposed Eigenray solutions. In this part we demonstrate the computational results for isotropic velocity fields. Note that in Appendices E, F and G of Part II we provide numerical/analytical comparisons between the Hamiltonian and the proposed Lagrangian approaches for different anisotropic symmetries (including spatially varying triclinic media). Two anisotropic numerical examples are included in Part VII, where both kinematic (stationary ray) and dynamic (geometric spreading) characteristics of the ray paths in ellipsoidal orthorhombic media have been computed.

In this study we mainly consider direct waves between two endpoints (*S* and *R*). The transmission/reflection of the rays across interfaces are beyond the scope of this study, but we briefly discuss them in Appendix I. We show that discontinuities in the ray direction can be naturally imposed at the assembly stage of the finite elements into the global structure, and we provide an example of the global traveltime gradient vector and Hessian matrix with this kind of direction discontinuity. The extensions with transmissions and reflections will be fully covered in our future work, together with other examples involving more realistic anisotropic elastic media.

In the following three examples, two-node Hermite elements were used.

Example 1. Eigenrays in high-velocity half-space under constant velocity layer ("head wave")

Consider a 1D velocity model whose vertical profile, gradient and second derivative are shown in Figure 2. The smoothed velocity model is depth dependent and described by,

$$v = v_o + \frac{\Delta v}{2}\left(1 + \tanh\frac{z - z_h}{\Delta z_h}\right) \quad , \tag{10}$$

where $z \equiv x_3$, $v_o = 2\,\text{km/s}$ is the velocity of the "homogeneous" layer above the half-space, and $\Delta v = 2\,\text{km/s}$ is the difference between the velocity of the half-space and that of the overlying layer; thus, the half-space velocity is $v_h = v_o + \Delta v = 4\,\text{km/s}$. Actually, neither the overlying layer nor the half-space is homogeneous, due to the transition zone. Parameter $z_h = 1.5\,\text{km}$ is the mid-level of the vertical transition zone, and parameter $\Delta z_h = 0.2\,\text{km}$ is the characteristic distance that shows the extent of the transition zone (which is approximately $5\Delta z_h$). The offset $h = 10$ km.

Figures 3a and 3b show the Eigenray solution with five and twenty finite elements, respectively, and they are almost identical. The red line is the initial guess, presented by an elliptic arc, defined by three parameters: the bottom point depth and the endpoints' offset and take-off angle. The bottom point depth is 2 km, which is approximately the lower end of the transition zone. In Figure 3a, lines of different colors show the segments of the path corresponding to different finite elements. In Figure 3b, the stationary ray path is shown by a black line.

Other initial paths, with over-estimated maximum depth, lead to the same stationary solution. In the next example we show that in the case of multi-arrivals, the final solution is sensitive to the initial guess.

Example 2. Eigenrays in a medium with low-velocity elliptic anomaly

Consider a constant background velocity with an elliptic anomaly region of a lower velocity, as shown in Figure 4a. The coordinates of the ellipse center are $c_1, c_3$, and the semi-axes of the ellipse are $a_1, a_3$. The velocity field is described by an analytic function,

$$v(x_1, x_3) = v_o + \frac{\Delta v}{2}(\tanh A - 1) \quad , \qquad (11)$$

where,

$$A = \frac{1}{s_e}\left[\frac{(x_1 - c_1)^2}{a_1^2} + \frac{(x_3 - c_3)^2}{a_3^2} - 1\right] \quad . \qquad (12)$$

The background velocity outside the ellipse is indicated by $v_o$, and $v_o - \Delta v$ is the anomalous low velocity inside the ellipse. Negative $\Delta v$ leads to anomalous high velocity inside the ellipse. Parameter $s_e$ is the smoothing scale: the smaller $s_e$ is, the sharper the velocity change. For infinitesimal $s_e$, the velocity function becomes discontinuous. We accept the following parameters,

$$\begin{aligned} v_o &= 5\,\text{km/s}, & \Delta v &= 3\,\text{km/s}, & c_1 &= 5\,\text{km}, & c_3 &= 3\,\text{km}, \\ a_1 &= 3\,\text{km}, & a_3 &= 2\,\text{km}, & s_e &= 0.2 & & \end{aligned} \quad (13)$$

The source is located at the subsurface point with zero horizontal coordinate and depth $d_o = 6\,\text{km}$, and the receiver is on the surface, with the one-way offset $h_s = 10\,\text{km}$. The absolute value of the velocity gradient is shown in Figure 4b. Three different initial guesses lead to three different solutions shown in Figure 5 by green, black and blue lines. The red ellipse is the contour of the velocity anomaly. The corresponding initial trajectories are shown by dashed lines of the same colors. The green and black lines are the "shallow" and "deep" solutions that bypass the anomaly from above and below, respectively. The blue line is the "transmission" solution that penetrates into the anomalous region. The initial path for the transmission solution is the straight line connecting the endpoints. The initial paths for the shallow and deep solutions are segments of a rectangle with rounded corners, whose parametric equation reads,

$$x_1(\tau_u) = a_e |\cos \tau_u|^{1/m} \text{sgn}(\cos \tau_u), \quad x_3(\tau_u) = b_e |\sin \tau_u|^{1/m} \text{sgn}(\sin \tau_u), \quad (14)$$

where $\tau_u$ is a running (flow) parameter (in radians), $a_e$ and $b_e$ are "semi-axes", and $m_u \geq 1$ is a real-number parameter ($m_u = 1$ leads to an ellipse; we applied $m_u = 5$). The initial ray direction components are,

$$r_1(\tau_u) = -\frac{a_e}{d_u}|\cos\tau_u|^{1/m_u}\sin\tau_u|\sin\tau_u| \quad , \quad r_3(\tau_u) = +\frac{b_e}{d_u}|\sin\tau_u|^{1/m_u}\cos\tau_u|\cos\tau_u| \quad , \tag{15}$$

where,

$$d_u = \sqrt{a_e^2\left(\cos^2\tau_u\right)^{1/m_u}\sin^4\tau_u + b_e^2\left(\sin^2\tau_u\right)^{1/m_u}\cos^4\tau_u} \quad . \tag{16}$$

For the problem we solve, $a_e = 10$ km and $b_e = 6$ km. For the full rectangle, $0 \le \tau_u \le 2\pi$ (four quadrants). One can consider a single quadrant for a shallow/deep initial guess.

For the "shallow" and "deep" rays (black and green lines), the resulting (minimum) traveltime is $t = 2.61048$ s, and for the "transmission" ray (blue line), the traveltime is $t = 3.71291$ s. "Shallow" and "deep" rays are symmetric solutions, where the ray bypasses the low-velocity inclusion and almost avoids penetration into the transition zone. These rays travel completely through the high-velocity background. For a "transmission" ray, refraction occurs twice, upon entry to and exit from the low-velocity ellipse. Thus, we deal with a multi-arrival case, characterized by three local minima, two of which are also global.

Example 3: Eigenrays in velocity field with two elliptic anomalies

Consider a model that combines slow- and high-velocity elliptic anomalies and a deep high-velocity half-space. It can be analytically described by,

$$v(x_1, x_3) = v_o - \frac{\Delta v}{2}(1 - \tanh A_a) + \frac{\Delta v}{2}(1 - \tanh A_b) + \frac{\Delta v_h}{2}(1 + \tanh A_c) \tag{17}$$

where

$$A_a = \frac{1}{s_e}\left[\frac{(x_1-c_a)^2}{a_1^2} + \frac{(x_3-c_3)^2}{a_3^2} - 1\right],$$

$$A_b = \frac{1}{s_e}\left[\frac{(x_1-c_b)^2}{a_1^2} + \frac{(x_3-c_3)^2}{a_3^2} - 1\right],$$

$$A_c \equiv \frac{x_3 - x_t}{\Delta x_o}. \tag{18}$$

Parameters $c_a$ and $c_b$ are horizontal coordinates of central points of elliptic anomalies, $c_3$ is their common vertical coordinate, $x_t$ is the floor depth of the high-velocity half-space. Parameters $s_e$ and $\Delta x_o$ govern the extent of the transition zones,

$$\begin{aligned}
&v_o = 4\,\text{km/s}, \quad \Delta v = 2\,\text{km/s}, \quad \Delta v_h = 3\,\text{km/s}, \quad c_a = 4\,\text{km}, \\
&c_b = 20\,\text{km}, \quad c_3 = 2\,\text{km}, \quad a_1 = 4\,\text{km}, \quad a_3 = 1\,\text{km}, \\
&x_t = 4\,\text{km}, \quad \Delta x_o = 0.2\,\text{km}, \quad s_e = 0.2,
\end{aligned} \tag{19}$$

and the offset $h = 22$ km. The velocity distribution and absolute value of the velocity gradient are shown in Figures 6a and 6b, respectively.

Figure 7 shows the Eigenray traveltime minimization results. Note that the scales in the horizontal and vertical directions are different (the horizontal space is much longer than shown). The gray and red ellipses show the contours of the low- and high-velocity anomalies, respectively. The mid-level of the transition zone between the layer with anomalies and high-velocity half-space is 4 km. The three dashed lines show initial paths, and the corresponding solid lines are stationary trajectories (local minima). All three solutions bypass the low-velocity anomaly and penetrate the high-velocity anomaly. Two initial guesses, shown by dashed black and dashed blue lines, represent elliptic arcs. The black line solution bypasses the low-velocity anomaly from the right. It has a shortest path, but not the least traveltime (not the global minimum), because this path does not reach the high-velocity half-space. The blue line solution bypasses the low-velocity anomaly from the left. Therefore, its path is longer, but it reaches the high-velocity half-space and thus arrives within a shorter time. The red line shows one more solution. Its initial guess was obtained with the control points shown by bold red points. There are five red points in total, but only three of them are internal/independent, the other two are endpoints of the trajectory (source and receiver). Using the five points, a standard cubic spline is created, shown by a dashed red line. The corresponding solution bypasses the low-velocity anomaly from the right and reaches the high-velocity half-space, but the portion of the path run in the half-space is shorter than that of the "blue" solution; therefore, its traveltime is longer than that of the "blue", but shorter than that of the "black". The real numbers in the legend of Figure 7 show the traveltimes of the three stationary paths in milliseconds.

## CONCLUSIONS

In this part we describe the implementation of our proposed ray bending algorithm, referred to as the Eigenray method, whose theoretical formulation is presented in Parts I and II of this study. The Eigenray method has been developed to solve two-point ray tracing problems in 3D smooth heterogeneous general anisotropic elastic media. It is based on the finite element method with the Hermite interpolation. Starting with an initial (non-stationary) discretized trajectory between two given endpoints, we construct a target function that includes two essential constraints. It is optimized (minimized) by computing corrections to the spatial locations and directions of the trajectory that yield a stationary traveltime solution. The first constraint is related to the element lengths governing the locations (distribution) of the nodes along the stationary ray, where a higher local curvature of the path leads to a denser grid. Without this constraint, the resolving equation set is under-defined. The second constraint is the normalization of the length of the ray (group) velocity direction at each node. Explicit expressions for the traveltime and its first and second derivatives allow the implementation of the Newton method for the optimization, where we search for a minimum traveltime and the target function includes explicitly the traveltime. In cases where the ray trajectory includes caustics, we search for a saddle point stationary solution using the gradient method where the target function includes the traveltime gradient squared. The target function that includes the traveltime gradient squared is also used in cases where the type of the stationary solution is not known ahead.

As a general strategy for ray tracing in complex subsurface geological media with complex wave phenomena and multi-arrivals (several stationary solutions between the two fixed endpoints), we suggest starting with the ray shooting method, and then use the proposed Eigenray method to obtain valid (acceptable) stationary rays between the source and the receivers located at the remaining shadow zones.

## ACKNOWLEDGEMENT

The authors are grateful to Emerson for the financial and technical support of this study and for the permission to publish its results. The gratitude is extended to Ivan Pšenčík, Einar Iversen, Michael Slawinski, Alexey Stovas, Vladimir Grechka, and our colleague Beth Orshalimy, whose valuable remarks helped to improve the content and style of this paper.

## APPENDIX A. HERMITE ELEMENTS

Our proposed anisotropic Eigenray method is based on imposing continuities of the locations and ray velocity directions at the ray trajectory nodes. The ray direction Cartesian components are derivatives of the corresponding location components wrt the arclength. Thus, the ray trajectory along the intervals is presented by the Hermite interpolation polynomials. The Hermite polynomials provide the interpolation of a function accounting for its nodal values and the nodal values of its derivatives (e.g., Hildebrand, 1987; Burden and Faires, 2005). The Hermite finite elements naturally support these conditions. In 3D space, each node has three location components and three direction components, but since the direction is described by a tangent vector of unit length, the number of independent DoF is five per node.

In this appendix, we follow conventional rules to construct the Hermite polynomials for two-node and three-node finite elements.

The internal unitless flow parameter

To keep the limits of integration fixed and identical for all finite elements, it is natural and convenient to define an internal unitless flow parameter, $\xi$, $-1 \leq \xi \leq 1$, inside the finite elements, to be used instead of the arclength *s*. For this we introduce a new, "normalized" Lagrangian, $\hat{L}(\xi)$, (see Table 2 in Part I) and formulate the governing relationships in terms of the internal parameter $\xi$, where the transformation metric relates the arclength *s* to $\xi$ (see Appendix C).

Two-node Hermite Element

An element with two end nodes has twelve DoF, only ten of which are independent due to the abovementioned constraints. Let *A* and *B* be the "left" and "right" ends of a finite element, as shown in Figure 8. The traveltime increases from *A* to *B*, $t_a < t_b$. The internal unitless parameter $\xi$ takes the values $-1$ and $+1$ at the element endpoints *A* and *B*, respectively.

Assume that a function $f(\xi)$ (that may be, for example, a scalar function or any Cartesian component of a vector) and its derivative, $f'(\xi) = df/d\xi$, are specified at the endpoints,

$$\{f_a, f'_a, f_b, f'_b\} \quad . \tag{A1}$$

The interpolation function can be presented as,

$$f(\xi) = f_a h_a(\xi) + f'_a d_a(\xi) + f_b h_b(\xi) + f'_b d_b(\xi) \quad . \tag{A2}$$

Functions $h_a(\xi), d_a(\xi), h_b(\xi), d_b(\xi)$ are all cubic polynomials, given by,

$$h_a(\xi) = +\frac{(1-\xi)^2(2+\xi)}{4} \quad , \quad d_a(\xi) = +\frac{(1-\xi)^2(1+\xi)}{4} \quad ,$$

$$h_b(\xi) = +\frac{(1+\xi)^2(2-\xi)}{4} \quad , \quad d_b(\xi) = -\frac{(1+\xi)^2(1-\xi)}{4} \quad . \tag{A3}$$

The shape functions are plotted in Figure 9.

Three-node Hermite Element

A three-node Hermite element, with the nodes at the endpoints $A, C$, $\xi = \pm 1$, and an additional central node $B$, $\xi = 0$, is shown in Figure 10. It provides better accuracy of the ray path for the same total number of trajectory nodes.

Given the nodal function values and nodal derivatives at the three nodes $A, B, C$,

$$\{f_a, f'_a, f_b, f'_b, f_c, f'_c\} \quad . \tag{A4}$$

The interpolation function reads,

$$f(\xi) = f_a h_a(\xi) + f_b h_b(\xi) + f_c h_c(\xi) + f'_a d_a(\xi) + f'_b d_b(\xi) + f'_c d_c(\xi) \quad , \tag{A5}$$

where the interpolation (shape) functions have properties similar to those of the two-node element. Functions $h_a(\xi), d_a(\xi), h_b(\xi), d_b(\xi), d_c(\xi)$ are fifth-degree polynomials, and $h_c(\xi)$ is a fourth-degree polynomial (as it has to be an even function), given by,

$$\begin{aligned}
h_a &= +\frac{1}{4}(1-\xi)^2 \xi^2 (4+3\xi) \quad, & d_a &= +\frac{1}{4}(1-\xi)^2 \xi^2 (1+\xi) \quad, \\
h_b &= +\frac{1}{4}(1+\xi)^2 \xi^2 (4-3\xi) \quad, & d_b &= -\frac{1}{4}(1+\xi)^2 \xi^2 (1-\xi) \quad, \\
h_c &= +\left(1-\xi^2\right)^2 & d_c &= +\xi\left(1-\xi^2\right)^2 \quad.
\end{aligned} \tag{A6}$$

The shape functions are plotted in Figure 11.

The location of a ray trajectory point between the nodes of a finite element is given by,

$$\mathbf{x}(\xi) = \sum_{I=1}^{n} \mathbf{x}_I h_I(\xi) + \sum_{I=1}^{n} \mathbf{x}'_I d_I(\xi) \quad, \qquad \mathbf{x}' = d\mathbf{x}/d\xi \quad, \tag{A7}$$

where $I$ is the index of the node and $n = 2,3$ is the number of the nodes of the finite element. $I$ stands for $a,b$ in two-node elements and for $a,b,c$ in three-node elements.

## APPENDIX B. NUMERICAL INTEGRATION

Two-node Hermite element

The integrals of the shape functions are,

$$\int_{-1}^{+1} h_a(\xi)d\xi = 1 \quad , \qquad \int_{-1}^{+1} d_a(\xi)d\xi = +\frac{1}{3} \quad ,$$
$$\int_{-1}^{+1} h_b(\xi)d\xi = 1 \quad , \qquad \int_{-1}^{+1} d_b(\xi)d\xi = -\frac{1}{3} \quad .$$
(B1)

This makes it possible to integrate numerically (approximately) any interpolated function $f(\xi)$, given its values and derivatives at the end nodes *A* and *B* of the segment (e.g., Hildebrand, 1987),

$$\int_{-1}^{+1} f(\xi)d\xi = f_a + f_b + \frac{f'_a - f'_b}{3} \quad .$$
(B2)

The sum of the weights is 2, because the length of the interval is 2, from $\xi = -1$ to $\xi = +1$. When the accuracy of the above quadrature does not suffice, we split the whole range of the flow parameter, $-1 \leq \xi \leq +1$, into a number of sub-intervals, with internal nodes. For a single sub-interval,

$$\int_{\xi_k}^{\xi_{k+1}} f(\xi)d\xi = \frac{\Delta h}{2}\left[ f(\xi_k) + f(\xi_{k+1}) + \frac{f'(\xi_k) - f'(\xi_{k+1})}{3}\frac{\Delta h}{2} \right] \quad , \quad \Delta h = \xi_{k+1} - \xi_k \quad .$$
(B3)

For the entire interval, the derivatives at the internal nodes cancel each other out, and only the derivatives at the endpoints remain.

$$\int_{\xi_{\text{ini}}}^{\xi_{\text{fin}}} f(\xi)d\xi = \frac{\Delta h}{2}\left[ f(\xi_o) + 2f(\xi_1) + 2f(\xi_2) + \ldots + 2f(\xi_{n-1}) + f(\xi_n) \right]$$
$$+ \frac{f'(\xi_o) - f'(\xi_n)}{12}\Delta h^2 \quad ,$$
(B4)

where for $n_s$ sub-intervals,

$$\Delta h = \frac{\xi_{\text{fin}} - \xi_{\text{ini}}}{n_s} \quad , \qquad \begin{aligned} \xi_o &= \xi_{\text{ini}} \quad , \\ \xi_k &= \xi_o + k\Delta h \quad , \end{aligned} \qquad \begin{aligned} \xi_n &= \xi_{\text{fin}} \quad , \\ k &= 0,1,2\ldots n_s \quad . \end{aligned} \tag{B5}$$

Three-node Hermite element

The integrals of the shape functions are,

$$\int_{-1}^{+1} h_a(\xi) d\xi = +\frac{7}{15} \quad , \qquad \int_{-1}^{+1} d_a(\xi) d\xi = +\frac{1}{15} \quad ,$$

$$\int_{-1}^{+1} h_b(\xi) d\xi = +\frac{16}{15} \quad , \qquad \int_{-1}^{+1} d_b(\xi) d\xi = 0 \quad , \tag{B6}$$

$$\int_{-1}^{+1} h_c(\xi) d\xi = +\frac{7}{15} \quad , \qquad \int_{-1}^{+1} d_c(\xi) d\xi = -\frac{1}{15} \quad .$$

The quadrature scheme becomes,

$$\int_{-1}^{+1} f(\xi) d\xi = \frac{7 f_a + 16 f_b + 7 f_c + f'_a - f'_c}{15} \quad . \tag{B7}$$

Again, the sum of the weights is 2. Note that the derivative at the central node $f'_b$ does not contribute to the integral because $d_b$ is an odd function, so that the integrals from $\xi = -1$ to $\xi = 0$ and from $\xi = 0$ to $\xi = +1$ cancel each other out.

To increase accuracy, we split the entire range into an even number $n$ of sub-intervals. The integral over two successive sub-intervals becomes,

$$\int_{\xi_{k-1}}^{\xi_{k+1}} f(\xi)d\xi = \frac{\Delta h}{15}\left[7f(\xi_{k-1})+16f(\xi_k)+7f(\xi_{k+1})+f'(\xi_{k-1})h-f'(\xi_{k+1})\Delta h\right] ,\quad\text{(B8)}$$

$$\text{where}\quad \Delta h = x_{k+1}-x_k = x_k - x_{k-1} .$$

For the entire interval, we obtain,

$$\int_{\xi_{ini}}^{\xi_{fin}} f(x)dx = \frac{\Delta h}{15}\left[7f(\xi_o)+16f(\xi_1)+14f(\xi_2)+16f(\xi_3)+\cdots\right.$$
$$\left.\cdots+14f(\xi_{n-2})+16f(\xi_{n-1})+7f(\xi_n)\right]+\frac{f'(\xi_o)-f'(\xi_n)}{15}\Delta h^2 . \quad\text{(B9)}$$

Equations B4 and B9 (along with equation B5) can be considered modifications of the trapezoid and Simpson rules, respectively. Items with the derivatives of the integrand at the endpoints of the whole interval increase the accuracy of the quadrature. For the modified Simpson scheme, the number of sub-intervals should be even. More accurate quadrature formulae based on the Hermite interpolation have been suggested by Ujević (2004), but these relationships require higher derivatives.

### APPENDIX C. NODAL DERIVATIVES OF THE RAY PATH COORDINATES

In each iteration of the optimization procedure, the ray trajectory is approximately known from the previous iteration. The locations and ray velocity directions at the nodes are specified, and interpolation is applied between the nodes. However, to interpolate the function, its nodal derivatives should be specified wrt the internal flow parameter $\xi$. Note that we only have the derivatives of the Cartesian coordinates wrt the arclength,

$$r_i = \frac{dx_i}{ds} \quad \text{or} \quad \mathbf{r} = \frac{d\mathbf{x}}{ds} \equiv \dot{\mathbf{x}} \quad , \quad i=1,2,3 \quad , \quad \mathbf{r}\cdot\mathbf{r}=1 \quad . \tag{C1}$$

To obtain the derivatives of the Cartesian components wrt parameter $\xi$, we apply the chain rule,

$$\frac{dx_i}{d\xi} = \frac{dx_i}{ds}\frac{ds}{d\xi} = r_i\, s' \quad \text{or} \quad \mathbf{x}'(\xi) = \mathbf{r}(\xi)s'(\xi) \quad . \tag{C2}$$

Thus, the missing values are the derivatives of the arclength $s'$ at the nodal points. Note that the locations $\mathbf{x}$ and the ray directions $\mathbf{r}$ are continuous at the end nodes of the adjacent elements. However, the derivatives $\mathbf{x}' = d\mathbf{x}/d\xi$ need not to be continuous: Jumps are allowed. Note that for the case of three-node elements, at the central nodes, both the ray direction $\dot{\mathbf{x}} = \mathbf{r} = d\mathbf{x}/ds$ and the derivative $\mathbf{x}' = d\mathbf{x}/d\xi$ are continuous.

In order to define the nodal derivatives of the arclength wrt the internal parameter, $s' = ds/d\xi$, an additional assumption or constraint is needed. We follow the method described by Yong and Cheng (2004) which is based on minimizing the pseudo strain energy of a finite-element ray path. In the cited paper, two-node intervals are considered, and the internal parameter range is $0 \leq \xi \leq 1$, while we use a symmetric interval $-1 \leq \xi \leq 1$, which makes the governing equations also symmetric. Therefore, we obtain the resulting equations in a different form, but the idea is the same: Assign such values to the nodal derivatives, $ds_I/d\xi$ (where $I$ is the node index), that lead to a minimum pseudo strain energy of the entire finite element. Assume that the finite element is an elastic rod with a constant cross-section. Then its specific strain energy (per unit length) is proportional to the curvature squared. The full pseudo strain energy is the integral of the specific energy within the segment length. The authors of the work cited above consider that the unsigned curvature $k$ of the line is given by the second derivative of the location,

$$\kappa(\xi) = \left| d^2\mathbf{x}/d\xi^2 \right| = \left| \mathbf{x}'' \right| \quad , \tag{C3}$$

and integrate the curvature squared over $\xi$ within the range $-1 \leq \xi \leq 1$. The pseudo strain energy of the finite element becomes,

$$E = \frac{1}{2} \int_{-1}^{+1} \kappa^2(\xi) d\xi \to \min \quad . \tag{C4}$$

This approach may be criticized. First, this second derivative is a principal factor of the curvature, but not its exact value, and second, the integration for the whole energy should be done over the arclength rather than over the internal parameter. Nevertheless, it leads to a clear and simple final relationship and we follow the suggested approximation. At the end of this appendix, we present the exact equation for the pseudo strain energy.

Two-node Hermite Element

The location of a trajectory point inside an element is given by equations A2 and A7, where the shape functions are provided by equation A3. Then the second derivative of a point location reads,

$$\mathbf{x}'' = \frac{\mathbf{x}'_b - \mathbf{x}'_a}{2} + 3\xi \frac{\mathbf{x}_a - \mathbf{x}_b + \mathbf{x}'_a + \mathbf{x}'_b}{2} \quad . \tag{C5}$$

The curvature squared represents a scalar product of a vector with itself,

$$\kappa^2 = \mathbf{x}'' \cdot \mathbf{x}'' \quad . \tag{C6}$$

The specific strain energy (per unit length) is assumed proportional to the curvature squared,

$$\frac{\kappa^2(\xi)}{2} = \frac{1}{2}\left(\frac{\mathbf{x}'_b - \mathbf{x}'_a}{2} + 3\xi\frac{\mathbf{x}_a - \mathbf{x}_b + \mathbf{x}'_a + \mathbf{x}'_b}{2}\right)^2 \qquad . \tag{C7}$$

The whole strain energy is the integral of the specific energy over the finite-element length,

$$E = \frac{1}{2}\int_{-1}^{+1}\kappa^2 d\xi = \mathbf{x}'^2_a + \mathbf{x}'_a \cdot \mathbf{x}'_b + \mathbf{x}'^2_b - \frac{3}{2}(\mathbf{x}'_b + \mathbf{x}'_a)\cdot(\mathbf{x}_b - \mathbf{x}_a) + \frac{3}{4}(\mathbf{x}_b - \mathbf{x}_a)^2 \qquad . \tag{C8}$$

It is suitable to introduce the location shift $\Delta\mathbf{x}$,

$$\Delta\mathbf{x} = \mathbf{x}_b - \mathbf{x}_a \qquad , \tag{C9}$$

where $\Delta\mathbf{x}$ is the chord vector connecting the end nodes of the element. The strain energy simplifies to,

$$E = \mathbf{x}'^2_a + \mathbf{x}'_a \cdot \mathbf{x}'_b + \mathbf{x}'^2_b - \frac{3}{2}(\mathbf{x}'_a + \mathbf{x}'_b)\cdot\Delta\mathbf{x} + \frac{3}{4}\Delta\mathbf{x}\cdot\Delta\mathbf{x} \qquad . \tag{C10}$$

Apply equation C2 to the endpoints,

$$\mathbf{x}'_a = s'_a \mathbf{r}_a \quad , \quad \mathbf{x}'_b = s'_b \mathbf{r}_b \qquad . \tag{C11}$$

Introduction of equation C11 into equation C10 leads to,

$$E = s'^2_a \mathbf{r}^2_a + s'_a s'_b \mathbf{r}_a \cdot \mathbf{r}_b + s'^2_b \mathbf{r}^2_b - \frac{3}{2}(s'_a \mathbf{r}_a + s'_b \mathbf{r}_b)\cdot\Delta\mathbf{x} + \frac{3}{4}\Delta\mathbf{x}\cdot\Delta\mathbf{x} \qquad . \tag{C12}$$

Recall that $\mathbf{r}_a$ and $\mathbf{r}_b$ are unit vectors (directions of the ray velocity at the endpoints of a finite element), so that $\mathbf{r}^2_a = 1$ and $\mathbf{r}^2_b = 1$. The strain energy simplifies to,

$$E = s_a'^2 + s_a' s_b' \mathbf{r}_a \cdot \mathbf{r}_b + s_b'^2 - \frac{3}{2}(s_a' \mathbf{r}_a + s_b' \mathbf{r}_b) \cdot \Delta \mathbf{x} + \frac{3}{4} \Delta \mathbf{x} \cdot \Delta \mathbf{x} \qquad (C13)$$

Minimization of the strain energy wrt the endpoint derivatives of the arclength,

$$\frac{\partial E}{\partial s_a'} = 0 \quad , \quad \frac{\partial E}{\partial s_b'} = 0 \qquad (C14)$$

leads to the following linear equation set,

$$2 s_a' + s_b' \mathbf{r}_a \cdot \mathbf{r}_b = \frac{3}{2} \mathbf{r}_a \cdot \Delta \mathbf{x}$$
$$2 s_b' + s_a' \mathbf{r}_a \cdot \mathbf{r}_b = \frac{3}{2} \mathbf{r}_b \cdot \Delta \mathbf{x} \qquad (C15)$$

or in a matrix form,

$$\begin{bmatrix} 2 & \mathbf{r}_a \cdot \mathbf{r}_b \\ \mathbf{r}_a \cdot \mathbf{r}_b & 2 \end{bmatrix} \cdot \begin{bmatrix} s_a' \\ s_b' \end{bmatrix} = \frac{3}{2} \begin{bmatrix} \mathbf{r}_a \cdot \Delta \mathbf{x} \\ \mathbf{r}_b \cdot \Delta \mathbf{x} \end{bmatrix} \qquad (C16)$$

The solution is,

$$s_a' = \frac{3}{2} \frac{2 \mathbf{r}_a - (\mathbf{r}_a \cdot \mathbf{r}_b) \mathbf{r}_b}{4 - (\mathbf{r}_a \cdot \mathbf{r}_b)^2} \cdot \Delta \mathbf{x}$$
$$s_b' = \frac{3}{2} \frac{2 \mathbf{r}_b - (\mathbf{r}_a \cdot \mathbf{r}_b) \mathbf{r}_a}{4 - (\mathbf{r}_a \cdot \mathbf{r}_b)^2} \cdot \Delta \mathbf{x} \qquad (C17)$$

Yong and Cheng (2004) emphasize that both endpoint derivatives $s'_a$ and $s'_b$ should be positive, otherwise the direction of the trajectory at the endpoints will be changed to the opposite direction (this is an unwanted effect). They analyze the direction preservation condition for a 2D case. We extend their approach to a 3D case. For this it is suitable to introduce the new notations, related to three angles, $\theta_a$, $\theta_b$, $\theta_e$,

$$\mathbf{r}_a \cdot \Delta \mathbf{x} = l_e \cos \theta_a \quad , \qquad \mathbf{r}_b \cdot \Delta \mathbf{x} = l_e \cos \theta_b \quad , \qquad \mathbf{r}_a \cdot \mathbf{r}_b = \cos \theta_e \quad , \tag{C18}$$

where $l_e = |\Delta \mathbf{x}|$ is the scalar distance between the endpoints of the element, $\theta_a$ is the angle between the chord $\Delta \mathbf{x}$ and the ray velocity at node $A$, $\theta_b$ is the angle between the chord and the ray velocity at node $B$, and $\theta_e$ is the angle between the ray velocities at the two endpoints. With these notations, the endpoint derivatives corresponding to the minimum strain energy become,

$$\begin{aligned} s'_a &= \frac{3 l_e}{2} \frac{2 \cos \theta_a - \cos \theta_b \cos \theta_e}{4 - \cos^2 \theta_e} \\ s'_b &= \frac{3 l_e}{2} \frac{2 \cos \theta_b - \cos \theta_a \cos \theta_e}{4 - \cos^2 \theta_e} \end{aligned} \tag{C19}$$

The denominator is always positive, so that the direction preserving conditions become,

$$2 \cos \theta_a > \cos \theta_b \cos \theta_e \quad , \qquad 2 \cos \theta_b > \cos \theta_a \cos \theta_e \quad . \tag{C20}$$

In most cases these conditions are satisfied. However, it is essential to check them for each element. There is no remedy if they are not satisfied.

Comment: Equation set C16 represents vanishing scalar products of the endpoint curvature by the endpoint direction, $\mathbf{x}'' \cdot \mathbf{x}'$ (at nodes $A$ and $B$, respectively). According to equation C5,

$$\mathbf{x}''_a = +\frac{3}{2} \Delta \mathbf{x} - 2 \mathbf{x}'_a - \mathbf{x}'_b \quad , \qquad \mathbf{x}''_b = -\frac{3}{2} \Delta \mathbf{x} + \mathbf{x}'_a + 2 \mathbf{x}'_b \quad . \tag{C21}$$

Note that the second derivative of the arclength wrt the internal flow parameter $\xi$ reads,

$$s' = \sqrt{\mathbf{x}' \cdot \mathbf{x}'} \quad , \qquad s'' = \frac{\mathbf{x}'' \cdot \mathbf{x}'}{\sqrt{\mathbf{x}' \cdot \mathbf{x}'}} \quad . \tag{C22}$$

Thus, the second derivative of the curvature wrt the arclength vanishes at the endpoints of the two-node element, provided the minimum strain energy conditions are satisfied.

Three-node Hermite Element

The interpolation is based on the values of the function and its derivatives at the end nodes, equations A5 and A7, where the shape functions are listed in equation A6. The derivatives of the arclength $s$ wrt the flow parameter $\xi$ have to be found at the end nodes $A, C$ and at the central node $B$. We assign such values of the nodal derivatives, $s'_a = ds_a/d\xi$, $s'_b = ds_b/d\xi$ and $s'_c = ds_c/d\xi$ that lead to a minimum strain energy of the entire finite element $ABC$. Introducing equation A5 into equations C4 and C6, we obtain the strain energy $E$. The minimum energy conditions are,

$$\frac{\partial E}{\partial s'_a} = 0 \quad , \quad \frac{\partial E}{\partial s'_b} = 0 \quad , \quad \frac{\partial E}{\partial s'_c} = 0 \quad . \tag{C23}$$

The strain energy is a quadratic function of $s'_a, s'_b, s'_c$. The expression for the strain energy includes these parameters up to products and squares (with linear and constant terms as well). There is a single minimum point, obtained from the linear set,

$$\begin{bmatrix} 332 & 320\mathbf{r}_a \cdot \mathbf{r}_b & 38\ \mathbf{r}_a \cdot \mathbf{r}_c \\ 320\mathbf{r}_a \cdot \mathbf{r}_b & 1280 & 320\mathbf{r}_b \cdot \mathbf{r}_c \\ 38\ \mathbf{r}_a \cdot \mathbf{r}_c & 320\mathbf{r}_b \cdot \mathbf{r}_c & 332 \end{bmatrix} \begin{bmatrix} s'_a \\ s'_b \\ s'_c \end{bmatrix} = - \begin{bmatrix} (569\mathbf{x}_a - 448\mathbf{x}_b - 121\mathbf{x}_c) \cdot \mathbf{r}_a \\ 960(\mathbf{x}_a - \mathbf{x}_c) \cdot \mathbf{r}_b \\ (121\mathbf{x}_a + 448\mathbf{x}_b - 569\mathbf{x}_c) \cdot \mathbf{r}_c \end{bmatrix} \quad . \tag{C24}$$

Solving this set, we obtain $s'_a, s'_b, s'_c$. This set may be also arranged as,

$$\mathbf{v}_a \cdot \mathbf{r}_a = 0 \quad , \qquad \mathbf{v}_b \cdot \mathbf{r}_b = 0 \quad , \qquad \mathbf{v}_c \cdot \mathbf{r}_c = 0 \quad , \tag{C25}$$

where,

$$\begin{bmatrix} \mathbf{v}_a \\ \mathbf{v}_b \\ \mathbf{v}_c \end{bmatrix} = \begin{bmatrix} 332 & 320 & 38 \\ 320 & 1280 & 320 \\ 38 & 320 & 332 \end{bmatrix} \begin{bmatrix} \mathbf{x}'_a \\ \mathbf{x}'_b \\ \mathbf{x}'_c \end{bmatrix} + \begin{bmatrix} 569 & -448 & -121 \\ 960 & 0 & -960 \\ 121 & 448 & -569 \end{bmatrix} \begin{bmatrix} \mathbf{x}_a \\ \mathbf{x}_b \\ \mathbf{x}_c \end{bmatrix} \quad . \tag{C26}$$

The second derivatives of the arclength $s$ wrt the flow parameter $\xi$ are nonzero at the nodes, unlike the endpoints of the two-nodal Hermite element.

Exact Strain Energy

As mentioned, the curvature is not equal to the absolute value of the second derivative of the ray point location. For a parametrically-defined space curve in three dimensions given in Cartesian coordinates as,

$$\mathbf{x}(\xi) = \begin{bmatrix} x_1(\xi) & x_2(\xi) & x_3(\xi) \end{bmatrix} \quad , \tag{C27}$$

the unsigned curvature is defined by,

$$\kappa = \frac{\sqrt{(x'_2 x''_3 - x'_3 x''_2)^2 + (x'_3 x''_1 - x'_1 x''_3)^2 + (x'_1 x''_2 - x'_2 x''_1)^2}}{(x'^2_1 + x'^2_2 + x'^2_3)^{3/2}} \quad . \tag{C28}$$

where prime denotes differentiation wrt parameter $\xi$. This can be expressed independently of the coordinate system by means of the formula,

$$\kappa = \frac{|\mathbf{x}' \times \mathbf{x}''|}{|\mathbf{x}'|^3} \quad . \tag{C29}$$

The strain energy becomes,

$$E = \frac{1}{2} \int_0^{\Delta s} \kappa^2(s) ds = \frac{1}{2} \int_{-1}^{+1} \kappa^2(\xi) \frac{ds}{d\xi} d\xi = \frac{1}{2} \int_{-1}^{+1} \frac{(\mathbf{x}' \times \mathbf{x}'')^2}{|\mathbf{x}'|^6} \frac{ds}{d\xi} d\xi \quad . \tag{C30}$$

where $\Delta s$ is the arclength of the given finite element. Next, we account for equation C2, and the pseudo strain energy reduces to,

$$E = \frac{1}{2} \int_{-1}^{+1} \frac{(\mathbf{x}' \times \mathbf{x}'')^2}{|\mathbf{x}'|^5} d\xi = \frac{1}{2} \int_{-1}^{+1} \frac{|\mathbf{x}'|^2 |\mathbf{x}''|^2 \sin^2(\mathbf{x}',\mathbf{x}'')}{|\mathbf{x}'|^5} d\xi = \frac{1}{2} \int_{-1}^{+1} \frac{|\mathbf{x}''|^2 \sin^2(\mathbf{x}',\mathbf{x}'')}{|\mathbf{x}'|^3} d\xi \quad . \tag{C31}$$

Yong and Cheng (2004) take into account only the first factor of the integrand: the curvature squared, $|\mathbf{x}''|^2$. The second factor,

$$\frac{\sin^2(\mathbf{x}',\mathbf{x}'')}{|\mathbf{x}'|^3} \quad , \tag{C32}$$

is ignored. Still, it is a good approximation, because in fact we do not compute the strain energy – we just need the conditions where the energy reaches the minimum. Also, exact minimum conditions are not needed as we do not work with real energy – it is just a criterion to define some reasonable values for the free parameters: The nodal derivatives of the arclength wrt the internal coordinate $\xi$.

An alternative approach to constructing the Hermite polynomials, based on the minimum curvature variation, has been suggested by Chi et al. (2005). With this method, the curvature and the pseudo strain energy are allowed to be arbitrary (not necessarily small), but the curvature variation (which is approximated by the third derivative of the location wrt the arclength) has to be minimized,

$$E_W = \frac{1}{2} \int_{-1}^{+1} g_e(\xi) d\xi \to \min \quad , \tag{C33}$$

where,

$$g_e = \left| d^3 \mathbf{x} / d\xi^3 \right| = \left| \mathbf{x}''' \right| \quad . \tag{C34}$$

In this work, we minimize the strain energy to establish the nodal derivatives of the arclength wrt the internal parameter, $s'_k = ds_k/d\xi$.

# APPENDIX D. CONNECTION BETWEEN GRADIENTS AND HESSIANS OF THE NORMALIZED AND ARCLENGTH-RELATED LGRANGIANS

It is convenient to formulate the finite-element operators with the normalized Lagrangian $\hat{L}(\xi)$, defined with the unitless flow variable within a single element, $-1 \leq \xi \leq +1$, as this parameterization allows keeping the limits of integration for each element fixed (regardless of the different element lengths). Note that for the first-degree homogeneous Lagrangian (wrt the ray direction vector), the traveltime integral is invariant under such re-parameterization (e.g., Bliss, 1916).

In this appendix, we relate the derivatives of the normalized Lagrangian to the corresponding derivatives of the arclength-related Lagrangian. Recall that the derivatives of the arclength-related Lagrangian are listed in equation set F2 of Part I. We will establish two first and four second derivatives,

$$\frac{\partial \hat{L}}{\partial \mathbf{x}} = \hat{L}_{\mathbf{x}} \quad , \quad \frac{\partial \hat{L}}{\partial \mathbf{x}'} = \hat{L}_{\mathbf{x}'} \quad ,$$

$$\frac{\partial^2 \hat{L}}{\partial \mathbf{x}^2} = \hat{L}_{\mathbf{xx}} \quad , \quad \frac{\partial^2 \hat{L}}{\partial \mathbf{x} \partial \mathbf{x}'} = \hat{L}_{\mathbf{xx}'} \quad , \tag{D1}$$

$$\frac{\partial^2 \hat{L}}{\partial \mathbf{x}' \partial \mathbf{x}} = \hat{L}_{\mathbf{x}'\mathbf{x}} \quad , \quad \frac{\partial^2 \hat{L}}{\partial \mathbf{x}'^2} = \hat{L}_{\mathbf{x}'\mathbf{x}'} \quad .$$

For that, we introduce the metric $ds/d\xi$, and the function $l(\xi)$, which is twice the metric. The normalized Lagrangian, $\hat{L}(\xi)$, is connected to the arclength-related Lagrangian through the metric,

$$\hat{L}(\xi) = \frac{dt}{d\xi} = \frac{l(\xi)}{2} \frac{1}{v_{\text{ray}}(\xi)} = \frac{l(\xi)}{2} L(s) \quad , \tag{D2}$$

where,

$$l(\xi) = 2\frac{ds(\xi)}{d\xi} = 2s'(\xi) = 2\sqrt{\mathbf{x}'(\xi) \cdot \mathbf{x}'(\xi)} \quad . \tag{D3}$$

The derivative of the location wrt the internal parameter, $\mathbf{x}' = d\mathbf{x}/d\xi$, is related to the corresponding derivative wrt the arclength,

$$\mathbf{x}'(\xi) = \frac{d\mathbf{x}}{d\xi} = \frac{d\mathbf{x}}{ds}\frac{ds}{d\xi} = \frac{ds}{d\xi}\mathbf{r}(\xi) = \frac{l(\xi)}{2}\mathbf{r}(\xi) \quad , \quad \mathbf{r}(\xi) = \frac{2}{l(\xi)}\mathbf{x}'(\xi) \quad . \tag{D4}$$

Compute the spatial and directional gradients of the normalized Lagrangian,

$$\hat{L}_{\mathbf{x}} = \frac{\partial \hat{L}}{\partial \mathbf{x}} = -\frac{l(\xi)}{2v_{\text{ray}}^2(\xi)} \frac{\partial v_{\text{ray}}}{\partial \mathbf{x}} + \frac{1}{2v_{\text{ray}}} \frac{\partial l(\xi)}{\partial \mathbf{x}} = \frac{l(\xi)}{2} L_{\mathbf{x}}(\xi) + \frac{1}{2v_{\text{ray}}} \frac{\partial l(\xi)}{\partial \mathbf{x}} \quad , \tag{D5}$$

and,

$$\hat{L}_{\mathbf{x}'} = \frac{\partial \hat{L}}{\partial \mathbf{x}'} = -\frac{l(\xi)}{2v_{\text{ray}}^2(\xi)} \frac{\partial v_{\text{ray}}}{\partial \mathbf{x}'} + \frac{1}{2v_{\text{ray}}} \frac{\partial l}{\partial \mathbf{x}'} \quad . \tag{D6}$$

According to equation D3, function $l(\xi)$ depends on the derivatives of the ray trajectory location, $\mathbf{x}'(\xi)$, and is independent of the location itself, $\mathbf{x}(\xi)$. Equation D5 then simplifies to,

$$\hat{L}_{\mathbf{x}} = -\frac{l(\xi)}{2v_{\text{ray}}^2(\xi)} \nabla_{\mathbf{x}} v_{\text{ray}} = \frac{l(\xi)}{2} L_{\mathbf{x}}(\xi) \quad . \tag{D7}$$

Note that,

$$\frac{\partial v_{\text{ray}}}{\partial \mathbf{x}'} = \frac{1}{s'(\xi)} \frac{\partial v_{\text{ray}}}{\partial \mathbf{r}} = \frac{2\nabla_{\mathbf{r}} v_{\text{ray}}}{l(\xi)} \quad , \tag{D8}$$

and,

$$\frac{\partial l}{\partial \mathbf{x}'} = \frac{2\mathbf{x}'}{\sqrt{\mathbf{x}' \cdot \mathbf{x}'}} = \frac{4\mathbf{x}'}{l(\xi)} \quad . \tag{D9}$$

Introduction of equations D8 and D9 into D6 leads to,

$$\hat{L}_{\mathbf{x}'} = -\frac{\nabla_{\mathbf{r}} v_{\text{ray}}}{v_{\text{ray}}^2(\xi)} + \frac{2\mathbf{x}'}{l(\xi) v_{\text{ray}}(\xi)} = L_{\mathbf{r}} \quad . \tag{D10}$$

Compute the location and directional Hessians of the normalized Lagrangian, taking into account equation D9 for $\partial l / \partial \mathbf{x}'$,

$$\hat{L}_{\mathbf{xx}} = \frac{\partial^2 \hat{L}}{\partial \mathbf{x}^2} = \frac{\partial}{\partial \mathbf{x}} \frac{\partial \hat{L}}{\partial \mathbf{x}} = -\frac{\partial}{\partial \mathbf{x}} \left[ \frac{l(\xi) \nabla_{\mathbf{x}} v_{\text{ray}}}{2 v_{\text{ray}}^2(\xi)} \right] =$$
$$+ \frac{l(\xi) \nabla_{\mathbf{x}} v_{\text{ray}} \otimes \nabla_{\mathbf{x}} v_{\text{ray}}}{v_{\text{ray}}^3(\xi)} - \frac{l(\xi) \nabla_{\mathbf{x}} \nabla_{\mathbf{x}} v_{\text{ray}}}{2 v_{\text{ray}}^2(\xi)} = \frac{l(\xi)}{2} L_{\mathbf{xx}}(\xi) \quad,$$
(D11)

$$\hat{L}_{\mathbf{x}'\mathbf{x}'} = \frac{\partial^2 \hat{L}}{\partial \mathbf{x}'^2} = \frac{\partial}{\partial \mathbf{x}'} \frac{\partial \hat{L}}{\partial \mathbf{x}'} =$$
$$-\frac{\partial}{\partial \mathbf{x}'} \left[ \frac{\nabla_{\mathbf{r}} v_{\text{ray}}}{v_{\text{ray}}^2(\xi)} - \frac{2\mathbf{x}'}{l(\xi) v_{\text{ray}}(\xi)} \right] = -\frac{\partial}{\partial \mathbf{x}'} \left[ \frac{l(\xi)}{2 v_{\text{ray}}^2(\xi)} \frac{\partial v_{\text{ray}}}{\partial \mathbf{x}'} - \frac{2\mathbf{x}'}{l(\xi) v_{\text{ray}}(\xi)} \right] =$$
$$+ \frac{4 \nabla_{\mathbf{r}} v_{\text{ray}} \otimes \nabla_{\mathbf{r}} v_{\text{ray}}}{l(\xi) v_{\text{ray}}^3(\xi)} - \frac{4 \nabla_{\mathbf{r}} v_{\text{ray}} \otimes \mathbf{x}'}{l^2(\xi) v_{\text{ray}}^2(\xi)} - \frac{4 \mathbf{x}' \otimes \nabla_{\mathbf{r}} v_{\text{ray}}}{l^2(\xi) v_{\text{ray}}^2(\xi)} - \frac{2 \nabla_{\mathbf{r}} \nabla_{\mathbf{r}} v_{\text{ray}}}{l(\xi) v_{\text{ray}}^2(\xi)}$$
$$- \frac{8 \mathbf{x}' \otimes \mathbf{x}'}{l^3(\xi) v_{\text{ray}}(\xi)} + \frac{2 \mathbf{I}}{l(\xi) v_{\text{ray}}(\xi)} = \frac{2}{l(\xi)} L_{\mathbf{rr}}(\xi) \quad.$$
(D12)

Compute the mixed Hessians of the normalized Lagrangian,

$$\hat{L}_{\mathbf{xx}'} = \frac{\partial^2 \hat{L}}{\partial \mathbf{xx}'} = \frac{\partial}{\partial \mathbf{x}'} \frac{\partial \hat{L}}{\partial \mathbf{x}} = -\frac{\partial}{\partial \mathbf{x}'} \left[ \frac{l(\xi) \nabla_{\mathbf{x}} v_{\text{ray}}}{2 v_{\text{ray}}^2(\xi)} \right] =$$
$$+ \frac{2 \nabla_{\mathbf{x}} v_{\text{ray}} \otimes \nabla_{\mathbf{r}} v_{\text{ray}}}{v_{\text{ray}}^3(\xi)} - \frac{2 \nabla_{\mathbf{x}} v_{\text{ray}} \otimes \mathbf{x}'}{l(\xi) v_{\text{ray}}^2(\xi)} - \frac{\nabla_{\mathbf{x}} \nabla_{\mathbf{r}} v_{\text{ray}}}{v_{\text{ray}}^2(\xi)} = L_{\mathbf{xr}}(\xi) \quad,$$
(D13)

$$\hat{L}_{\mathbf{x}'\mathbf{x}} = \frac{\partial^2 \hat{L}}{\partial \mathbf{x}' \mathbf{x}} = \frac{\partial}{\partial \mathbf{x}} \frac{\partial \hat{L}}{\partial \mathbf{x}'} = -\frac{\partial}{\partial \mathbf{x}} \left[ \frac{\nabla_{\mathbf{r}} v_{\text{ray}}}{v_{\text{ray}}^2(\xi)} - \frac{2\mathbf{x}'}{l(\xi) v_{\text{ray}}(\xi)} \right] =$$
$$+ \frac{2 \nabla_{\mathbf{r}} v_{\text{ray}} \otimes \nabla_{\mathbf{x}} v_{\text{ray}}}{v_{\text{ray}}^3(\xi)} - \frac{2 \mathbf{x}' \otimes \nabla_{\mathbf{x}} v_{\text{ray}}}{l(\xi) v_{\text{ray}}^2(\xi)} - \frac{\nabla_{\mathbf{r}} \nabla_{\mathbf{x}} v_{\text{ray}}}{v_{\text{ray}}^2(\xi)} = L_{\mathbf{rx}}(\xi) \quad.$$
(D14)

We summarize the relationships between the gradients and Hessians of the normalized and arclength-related Lagrangians,

$$\hat{L}(\xi) = \frac{l(\xi)}{2} L(\xi) \quad , \quad \mathbf{r}(\xi) = \frac{2}{l(\xi)} \mathbf{x}'(\xi) \quad ,$$

$$\hat{L}_{\mathbf{x}}(\xi) = \frac{l(\xi)}{2} L_{\mathbf{x}}(\xi) \quad , \quad \hat{L}_{\mathbf{x}'}(\xi) = L_{\mathbf{x}}(\xi) \quad ,$$

$$\hat{L}_{\mathbf{xx}}(\xi) = \frac{l(\xi)}{2} L_{\mathbf{xx}}(\xi) \quad , \quad \hat{L}_{\mathbf{xr}}(\xi) = L_{\mathbf{xr}}(\xi) \quad , \tag{D15}$$

$$\hat{L}_{\mathbf{rx}}(\xi) = L_{\mathbf{rx}}(\xi) \quad , \quad \hat{L}_{\mathbf{rr}}(\xi) = \frac{2}{l(\xi)} L_{\mathbf{rx}}(\xi) \quad ,$$

$$l(\xi) = 2 ds / d\xi = 2 \sqrt{\mathbf{x}'(\xi) \cdot \mathbf{x}'(\xi)} \quad (2 \times \text{metric}) \quad .$$

## APPENDIX E. LOCAL TRAVELTIME

Applying numerical integration, we compute the traveltime within the element. It follows from the definition of the normalized Lagrangian, $\hat{L}(\xi)$, given in equations D2 and D3,

$$\Delta t = \int_{\xi=-1}^{\xi=+1} \frac{\sqrt{\mathbf{x}'(\xi) \cdot \mathbf{x}'(\xi)}}{v_{\text{ray}}[\mathbf{x}(\xi), \mathbf{r}(\xi)]} \frac{d\xi}{ds} ds = \int_{\xi=-1}^{\xi=+1} \frac{\sqrt{\mathbf{x}'(\xi) \cdot \mathbf{x}'(\xi)}}{v_{\text{ray}}[\mathbf{x}(\xi), \mathbf{x}'(\xi)]} d\xi = \int_{\xi=-1}^{\xi=+1} \hat{L}(\xi) d\xi \quad . \tag{E1}$$

Since $\mathbf{x}(\xi)$ is a polynomial, $\mathbf{x}'(\xi)$ and $\mathbf{x}''(\xi)$ can be easily computed, and the ray velocity along the element trajectory depends on the internal parameter,

$$v_{\text{ray}}(\xi) = v_{\text{ray}}[\mathbf{x}(\xi), \mathbf{x}'(\xi)] \quad . \tag{E2}$$

The element (local) traveltime becomes,

$$\Delta t = \int\limits_{\xi=-1}^{\xi=+1} \hat{L}(\xi) d\xi = \frac{1}{2} \int\limits_{\xi=-1}^{\xi=+1} \frac{l(\xi) d\xi}{v_{\text{ray}}(\xi)} \quad , \tag{E4}$$

and the element (local) arclength is the average value of $l(\xi)$ within the element,

$$\Delta s = \frac{1}{2} \int\limits_{\xi=-1}^{\xi=+1} l(\xi) d\xi \quad , \tag{E5}$$

that is the reason for choosing factor 2 in equation D2.

In order to improve the accuracy of the numerical element traveltime integral E4, we need to compute the numerical derivatives of the traveltime integrand wrt the internal flow parameter, $d\hat{L}(\xi)/d\xi$, at the endpoints of the finite element,

$$\frac{d\hat{L}(\xi)}{d\xi} = \frac{\partial \hat{L}}{\partial \mathbf{x}} \cdot \frac{\partial \mathbf{x}}{\partial \xi} + \frac{\partial \hat{L}}{\partial \mathbf{x}'} \cdot \frac{\partial \mathbf{x}'}{\partial \xi} = \hat{L}_{\mathbf{x}} \cdot \mathbf{x}'(\xi) + \hat{L}_{\mathbf{x}'} \cdot \mathbf{x}''(\xi) \quad , \tag{E6}$$

where the relationship between $\mathbf{x}'(\xi)$ and $\mathbf{r}(\xi)$ is provided in equation D4.

Introduction of equation set D15 into equation E6 results in,

$$\frac{d\hat{L}(\xi)}{d\xi} = \frac{l(\xi)}{2} L_{\mathbf{x}}(\xi) \cdot \mathbf{x}'(\xi) + L_{\mathbf{r}}(\xi) \cdot \mathbf{x}''(\xi) \quad . \tag{E7}$$

Each term on the right-hand side of equation E7 is a scalar value.

The formulae for $d\hat{L}/d\xi$ will be used at the ends of the interval only, i.e., for $\xi = \pm 1$. The ray trajectory and its derivatives wrt the flow parameter $\xi$ are given by,

$$\mathbf{x}(\xi) = \sum_{I=1}^{n} \mathbf{x}_I h_I(\xi) + \sum_{I=1}^{n} \mathbf{x}'_I d_I(\xi) ,$$

$$\mathbf{x}'(\xi) = \sum_{I=1}^{n} \mathbf{x}_I h'_I(\xi) + \sum_{I=1}^{n} \mathbf{x}'_I d'_I(\xi) , \quad \text{(E8)}$$

$$\mathbf{x}''(\xi) = \sum_{I=1}^{n} \mathbf{x}_I h''_I(\xi) + \sum_{I=1}^{n} \mathbf{x}'_I d''_I(\xi) ,$$

where $I$ is the successive node index within a single element (to be replaced by indices $a, b, c$), and $n$ is the number of element nodes. The shape (interpolation) functions $h_I(\xi)$ and $d_I(\xi)$ are listed in equations A3 and A6, for $n = 2$ and $n = 3$, respectively.

### APPENDIX F. SUB-BLOCKS OF LOCAL TRAVELTIME GRADIENT AND HESSIAN

The local gradient vector consists of $n$ sequential blocks, where $n$ is the number of nodes within an element. In this work we provide generic solutions that can be applied to two-node elements ($n = 2$) and three-node elements ($n = 3$). Each block includes two sub-blocks with spatial and directional traveltime derivatives, respectively, where each sub-block is of length 3 for 3D space. The length of the gradient block is 6, and the length of the local gradient is 12 or 18 for two-nodal and three-nodal elements, respectively. For a three-node element, the scheme of the local gradient is presented in Table 1, and the scheme of a single gradient block is presented in Table 2, where subscript $I$ substitutes for the successive nodal index $a, b, c$.

The local traveltime Hessian matrix consists of $n \times n$ blocks. Each block has 4 sub-blocks: spatial, directional, and two types of mixed sub-blocks, where derivatives are taken first wrt location, then wrt direction, or vice versa. Note that the two mixed sub-blocks are not identical. The mixed sub-blocks of the diagonal blocks are transposed wrt each other, and the mixed sub-blocks of the non-diagonal blocks are just different. Note that scalar mixed derivatives are independent of the order of variables, but this is not so in our case, where each of the traveltime Hessian sub-blocks is a matrix (tensor) of dimension $3 \times 3$. The dimensions of a single traveltime Hessian block are $6 \times 6$, and the dimensions of the local traveltime Hessian are $12 \times 12$ or $18 \times 18$ for two-nodal and three-nodal elements, respectively. The scheme of the local traveltime Hessian is presented in Table 3, and the scheme of a single traveltime Hessian block is presented in Tables 4 and 5, where subscripts $I, J$ substitute for the nodal indices $a, b$ or $a, b, c$. Symbol $\Delta$ emphasizes that we consider a local time, related to a single element, and its derivatives. The sub-blocks of the local traveltime gradient and Hessian derived in this appendix are generic and valid for both two-node and three-node elements. Only the shape (interpolation) functions differ.

The local traveltime is defined by equation E1, and the sub-blocks of the local traveltime gradient and Hessian become,

$$\frac{\partial \Delta t}{\partial \mathbf{x}_I} = \frac{\partial}{\partial \mathbf{x}_I} \int_{-1}^{+1} \hat{L}(\xi) d\xi = \int_{-1}^{+1} \frac{\partial \hat{L}(\xi)}{\partial \mathbf{x}_I} d\xi \quad \text{spatial gradient} \quad , \tag{F1}$$

$$\frac{\partial \Delta t}{\partial \mathbf{x}'_I} = \frac{\partial}{\partial \mathbf{x}'_I} \int_{-1}^{+1} \hat{L}(\xi) d\xi = \int_{-1}^{+1} \frac{\partial \hat{L}(\xi)}{\partial \mathbf{x}'_I} d\xi \quad \text{directional gradient} \quad , \tag{F2}$$

$$\frac{\partial^2 \Delta t}{\partial \mathbf{x}_I \partial \mathbf{x}_J} = \frac{\partial^2}{\partial \mathbf{x}_I \partial \mathbf{x}_J} \int_{-1}^{+1} \hat{L}(\xi) d\xi = \int_{-1}^{+1} \frac{\partial^2 \hat{L}(\xi)}{\partial \mathbf{x}_I \partial \mathbf{x}_J} d\xi \quad \text{spatial Hessian} \quad , \tag{F3}$$

$$\frac{\partial^2 \Delta t}{\partial \mathbf{x}'_I \partial \mathbf{x}'_J} = \frac{\partial^2}{\partial \mathbf{x}'_I \partial \mathbf{x}'_J} \int_{-1}^{+1} \hat{L}(\xi) d\xi = \int_{-1}^{+1} \frac{\partial^2 \hat{L}(\xi)}{\partial \mathbf{x}'_I \partial \mathbf{x}'_J} d\xi \quad \text{directional Hessian} \quad , \tag{F4}$$

$$\frac{\partial^2 \Delta t}{\partial \mathbf{x}_I \partial \mathbf{x}'_J} = \frac{\partial^2}{\partial \mathbf{x}_I \partial \mathbf{x}'_J} \int_{-1}^{+1} \hat{L}(\xi) d\xi = \int_{-1}^{+1} \frac{\partial^2 \hat{L}(\xi)}{\partial \mathbf{x}_I \partial \mathbf{x}'_J} d\xi \qquad \text{mixed } \mathbf{xx}' \text{ Hessian} \quad , \tag{F5}$$

$$\frac{\partial^2 \Delta t}{\partial \mathbf{x}'_I \partial \mathbf{x}_J} = \frac{\partial^2}{\partial \mathbf{x}'_I \partial \mathbf{x}_J} \int_{-1}^{+1} \hat{L}(\xi) d\xi = \int_{-1}^{+1} \frac{\partial^2 \hat{L}(\xi)}{\partial \mathbf{x}'_I \partial \mathbf{x}_J} d\xi \qquad \text{mixed } \mathbf{x}'\mathbf{x} \text{ Hessian} \quad . \tag{F6}$$

Since the limits of the integration are independent of $\mathbf{x}_I$ and $\mathbf{x}_J$, we can switch the differential and integral operators.

Thus, what we need are two first derivatives (vectors of length 3) and four second derivatives (tensors of dimension $3 \times 3$),

$$\frac{\partial \hat{L}}{\partial \mathbf{x}_I} \quad , \quad \frac{\partial \hat{L}}{\partial \mathbf{x}'_I} \quad , \quad \frac{\partial^2 \hat{L}}{\partial \mathbf{x}_I \partial \mathbf{x}_J} \quad , \quad \frac{\partial^2 \hat{L}}{\partial \mathbf{x}'_I \partial \mathbf{x}'_J} \quad , \quad \frac{\partial^2 \hat{L}}{\partial \mathbf{x}_I \partial \mathbf{x}'_J} \quad , \quad \frac{\partial^2 \hat{L}}{\partial \mathbf{x}'_I \partial \mathbf{x}_J} \quad . \tag{F7}$$

For the spatial and directional sub-blocks of the local gradient, we obtain (with equation set D15),

$$\frac{\partial \hat{L}}{\partial \mathbf{x}_I} = \frac{\partial \hat{L}}{\partial \mathbf{x}} \frac{\partial \mathbf{x}}{\partial \mathbf{x}_I} + \frac{\partial \hat{L}}{\partial \mathbf{x}'} \frac{\partial \mathbf{x}'}{\partial \mathbf{x}_I} = h_I(\xi) \hat{L}_\mathbf{x} + h'_I(\xi) \hat{L}_{\mathbf{x}'} = \frac{l(\xi)}{2} h_I(\xi) L_\mathbf{x} + h'_I(\xi) L_\mathbf{r} \quad , \tag{F8}$$

and,

$$\frac{\partial \hat{L}}{\partial \mathbf{x}'_I} = \frac{\partial \hat{L}}{\partial \mathbf{x}} \frac{\partial \mathbf{x}}{\partial \mathbf{x}'_I} + \frac{\partial \hat{L}}{\partial \mathbf{x}'} \frac{\partial \mathbf{x}'}{\partial \mathbf{x}'_I} = d_I(\xi) \hat{L}_\mathbf{x} + d'_I(\xi) \hat{L}_{\mathbf{x}'} = \frac{l(\xi)}{2} d_I(\xi) L_\mathbf{x} + d'_I(\xi) L_\mathbf{r} \quad , \tag{F9}$$

where $I$ is the block number.

For the spatial and directional sub-blocks of the local Hessian, we obtain,

$$\frac{\partial^2 \hat{L}}{\partial \mathbf{x}_I \partial \mathbf{x}_J} = \frac{\partial}{\partial \mathbf{x}} \left[ h_I(\xi) \frac{\partial \hat{L}}{\partial \mathbf{x}} + h'_I(\xi) \frac{\partial \hat{L}}{\partial \mathbf{x}'} \right] \frac{\partial \mathbf{x}}{\partial \mathbf{x}_J} + \frac{\partial}{\partial \mathbf{x}'} \left[ h_I(\xi) \frac{\partial \hat{L}}{\partial \mathbf{x}} + h'_I(\xi) \frac{\partial \hat{L}}{\partial \mathbf{x}'} \right] \frac{\partial \mathbf{x}'}{\partial \mathbf{x}_J}$$
$$= h_I(\xi) h_J(\xi) \frac{\partial^2 \hat{L}}{\partial \mathbf{x}^2} + h_I(\xi) h'_J(\xi) \frac{\partial^2 \hat{L}}{\partial \mathbf{x} \partial \mathbf{x}'} + h'_I(\xi) h_J(\xi) \frac{\partial^2 \hat{L}}{\partial \mathbf{x}' \partial \mathbf{x}} + h'_I(\xi) h'_J(\xi) \frac{\partial^2 \hat{L}}{\partial \mathbf{x}'^2} \; ,$$
(F10)

and,

$$\frac{\partial^2 \hat{L}}{\partial \mathbf{x}'_I \partial \mathbf{x}'_J} = \frac{\partial}{\partial \mathbf{x}} \left[ d_I(\xi) \frac{\partial \hat{L}}{\partial \mathbf{x}} + d'_I(\xi) \frac{\partial \hat{L}}{\partial \mathbf{x}'} \right] \frac{\partial \mathbf{x}}{\partial \mathbf{x}'_J} + \frac{\partial}{\partial \mathbf{x}'} \left[ d_I(\xi) \frac{\partial \hat{L}}{\partial \mathbf{x}} + d'_I(\xi) \frac{\partial \hat{L}}{\partial \mathbf{x}'} \right] \frac{\partial \mathbf{x}'}{\partial \mathbf{x}'_J}$$
$$= d_I(\xi) d_J(\xi) \frac{\partial^2 \hat{L}}{\partial \mathbf{x}^2} + d_I(\xi) d'_J(\xi) \frac{\partial^2 \hat{L}}{\partial \mathbf{x} \partial \mathbf{x}'} + d'_I(\xi) d_J(\xi) \frac{\partial^2 \hat{L}}{\partial \mathbf{x}' \partial \mathbf{x}} + d'_I(\xi) d'_J(\xi) \frac{\partial^2 \hat{L}}{\partial \mathbf{x}'^2} \; .$$
(F11)

For the two mixed sub-blocks of the local Hessian, we obtain,

$$\frac{\partial^2 \hat{L}}{\partial \mathbf{x}_I \partial \mathbf{x}'_J} = \frac{\partial}{\partial \mathbf{x}} \left[ h_I(\xi) \frac{\partial \hat{L}}{\partial \mathbf{x}} + h'_I(\xi) \frac{\partial \hat{L}}{\partial \mathbf{x}'} \right] \frac{\partial \mathbf{x}}{\partial \mathbf{x}'_J} + \frac{\partial}{\partial \mathbf{x}'} \left[ h_I(\xi) \frac{\partial \hat{L}}{\partial \mathbf{x}} + h'_I(\xi) \frac{\partial \hat{L}}{\partial \mathbf{x}'} \right] \frac{\partial \mathbf{x}'}{\partial \mathbf{x}'_J}$$
$$= h_I(\xi) d_J(\xi) \frac{\partial^2 \hat{L}}{\partial \mathbf{x}^2} + h_I(\xi) d'_J(\xi) \frac{\partial^2 \hat{L}}{\partial \mathbf{x} \partial \mathbf{x}'} + h'_I(\xi) d_J(\xi) \frac{\partial^2 \hat{L}}{\partial \mathbf{x}' \partial \mathbf{x}} + d'_I(\xi) d'_J(\xi) \frac{\partial^2 \hat{L}}{\partial \mathbf{x}'^2} \; ,$$
(F12)

and

$$\frac{\partial^2 \hat{L}}{\partial \mathbf{x}'_I \partial \mathbf{x}_J} = \frac{\partial}{\partial \mathbf{x}} \left[ d_I(\xi) \frac{\partial \hat{L}}{\partial \mathbf{x}} + d'_I(\xi) \frac{\partial \hat{L}}{\partial \mathbf{x}'} \right] \frac{\partial \mathbf{x}}{\partial \mathbf{x}_J} + \frac{\partial}{\partial \mathbf{x}'} \left[ d_I(\xi) \frac{\partial \hat{L}}{\partial \mathbf{x}} + d'_I(\xi) \frac{\partial \hat{L}}{\partial \mathbf{x}'} \right] \frac{\partial \mathbf{x}'}{\partial \mathbf{x}_J}$$
$$= d_I(\xi) h_J(\xi) \frac{\partial^2 \hat{L}}{\partial \mathbf{x}^2} + d_I(\xi) h'_J(\xi) \frac{\partial^2 \hat{L}}{\partial \mathbf{x} \partial \mathbf{x}'} + d'_I(\xi) h_J(\xi) \frac{\partial^2 \hat{L}}{\partial \mathbf{x}' \partial \mathbf{x}} + d'_I(\xi) h'_J(\xi) \frac{\partial^2 \hat{L}}{\partial \mathbf{x}'^2} \; .$$
(F13)

To compute the quadrature accurately, we will need the derivatives of the Lagrangian gradient integrand wrt the internal flow parameter $\xi$, at the ends of the interval, $\xi = \pm 1$.

$$\frac{d}{d\xi} \frac{\partial \hat{L}}{\partial \mathbf{x}_I} \quad \text{and} \quad \frac{d}{d\xi} \frac{\partial \hat{L}}{\partial \mathbf{x}'_I} \quad .$$
(F14)

Applying the chain rule, we obtain,

$$\frac{d}{d\xi}\frac{\partial \hat{L}}{\partial \mathbf{x}_I} = \frac{\partial}{\partial \xi}\frac{\partial \hat{L}}{\partial \mathbf{x}_I} + \frac{\partial}{\partial \mathbf{x}}\frac{\partial \hat{L}}{\partial \mathbf{x}_I}\cdot\frac{d\mathbf{x}}{d\xi} + \frac{\partial}{\partial \mathbf{x}'}\frac{\partial \hat{L}}{\partial \mathbf{x}_I}\cdot\frac{d\mathbf{x}'}{d\xi} =$$
$$\frac{\partial}{\partial \xi}\left[\frac{\partial \hat{L}}{\partial \mathbf{x}}\frac{d\mathbf{x}}{d\mathbf{x}_I} + \frac{\partial \hat{L}}{\partial \mathbf{x}'}\frac{d\mathbf{x}'}{d\mathbf{x}_I}\right] + \frac{\partial}{\partial \mathbf{x}}\left[\frac{\partial \hat{L}}{\partial \mathbf{x}}\frac{d\mathbf{x}}{d\mathbf{x}_I} + \frac{\partial \hat{L}}{\partial \mathbf{x}'}\frac{d\mathbf{x}'}{d\mathbf{x}_I}\right]\cdot\mathbf{x}' + \frac{\partial}{\partial \mathbf{x}'}\left[\frac{\partial \hat{L}}{\partial \mathbf{x}}\frac{d\mathbf{x}}{d\mathbf{x}_I} + \frac{\partial \hat{L}}{\partial \mathbf{x}'}\frac{d\mathbf{x}'}{d\mathbf{x}_I}\right]\cdot\mathbf{x}''$$
$$= h_I'(\xi)\hat{L}_\mathbf{x} + h_I''(\xi)\hat{L}_{\mathbf{x}'}$$
$$+ h_I(\xi)\frac{\partial^2 \hat{L}}{\partial \mathbf{x}^2}\cdot\mathbf{x}' + h_I'(\xi)\frac{\partial^2 \hat{L}}{\partial \mathbf{x}'\partial \mathbf{x}}\cdot\mathbf{x}' + h_I(\xi)\frac{\partial^2 \hat{L}}{\partial \mathbf{x}\partial \mathbf{x}'}\cdot\mathbf{x}'' + h_I'(\xi)\frac{\partial^2 \hat{L}}{\partial \mathbf{x}'^2}\cdot\mathbf{x}'' \quad, \tag{F15}$$

and

$$\frac{d}{d\xi}\frac{\partial \hat{L}}{\partial \mathbf{x}_I'} = \frac{\partial}{\partial \xi}\frac{\partial \hat{L}}{\partial \mathbf{x}_I'} + \frac{\partial}{\partial \mathbf{x}}\frac{\partial \hat{L}}{\partial \mathbf{x}_I'}\cdot\frac{d\mathbf{x}}{d\xi} + \frac{\partial}{\partial \mathbf{x}'}\frac{\partial \hat{L}}{\partial \mathbf{x}_I'}\cdot\frac{d\mathbf{x}'}{d\xi} =$$
$$\frac{\partial}{\partial \xi}\left[\frac{\partial \hat{L}}{\partial \mathbf{x}}\frac{d\mathbf{x}}{d\mathbf{x}_I'} + \frac{\partial \hat{L}}{\partial \mathbf{x}'}\frac{d\mathbf{x}'}{d\mathbf{x}_I'}\right] + \frac{\partial}{\partial \mathbf{x}}\left[\frac{\partial \hat{L}}{\partial \mathbf{x}}\frac{d\mathbf{x}}{d\mathbf{x}_I'} + \frac{\partial \hat{L}}{\partial \mathbf{x}'}\frac{d\mathbf{x}'}{d\mathbf{x}_I'}\right]\cdot\mathbf{x}' + \frac{\partial}{\partial \mathbf{x}'}\left[\frac{\partial \hat{L}}{\partial \mathbf{x}}\frac{d\mathbf{x}}{d\mathbf{x}_I'} + \frac{\partial \hat{L}}{\partial \mathbf{x}'}\frac{d\mathbf{x}'}{d\mathbf{x}_I'}\right]\cdot\mathbf{x}''$$
$$= d_I'(\xi)\hat{L}_\mathbf{x} + d_I''(\xi)\hat{L}_{\mathbf{x}'}$$
$$+ d_I(\xi)\frac{\partial^2 \hat{L}}{\partial \mathbf{x}^2}\cdot\mathbf{x}' + d_I'(\xi)\frac{\partial^2 \hat{L}}{\partial \mathbf{x}'\partial \mathbf{x}}\cdot\mathbf{x}' + d_I(\xi)\frac{\partial^2 \hat{L}}{\partial \mathbf{x}\partial \mathbf{x}'}\cdot\mathbf{x}'' + d_I'(\xi)\frac{\partial^2 \hat{L}}{\partial \mathbf{x}'^2}\cdot\mathbf{x}'' \quad. \tag{F16}$$

Introduction of equation set D15 into equations F15 and F16 leads to,

$$\frac{d}{d\xi}\frac{\partial \hat{L}}{\partial \mathbf{x}_I} = +\frac{l(\xi)h_I'(\xi)}{2}L_\mathbf{x} + h_I''(\xi)L_\mathbf{r}$$
$$+\frac{l(\xi)h_I(\xi)}{2}L_{\mathbf{xx}}\cdot\mathbf{x}' + h_I'(\xi)L_{\mathbf{rx}}\cdot\mathbf{x}' + h_I(\xi)L_{\mathbf{xr}}\cdot\mathbf{x}'' + \frac{2h_I'(\xi)}{l(\xi)}L_{\mathbf{rr}}\cdot\mathbf{x}'' \quad, \tag{F17}$$

$$\frac{d}{d\xi}\frac{\partial \hat{L}}{\partial \mathbf{x}_I'} = \frac{l(\xi)d_I'(\xi)}{2}L_\mathbf{x} + d_I''(\xi)L_\mathbf{r}$$
$$+\frac{l(\xi)d_I(\xi)}{2}L_{\mathbf{xx}}\cdot\mathbf{x}' + d_I'(\xi)L_{\mathbf{rx}}\cdot\mathbf{x}' + d_I(\xi)L_{\mathbf{xr}}\cdot\mathbf{x}'' + \frac{2d_I'(\xi)}{l(\xi)}L_{\mathbf{rr}}\cdot\mathbf{x}'' \quad. \tag{F18}$$

To compute the sub-blocks of the local Hessian, we introduce equation set D15 into equations F10-F13,

$$\frac{\partial^2 \hat{L}}{\partial \mathbf{x}_I \partial \mathbf{x}_J} = \frac{l(\xi) h_I(\xi) h_J(\xi)}{2} L_{\mathbf{xx}}(\xi) + h_I(\xi) h'_J(\xi) L_{\mathbf{xr}}(\xi) \\ + h'_I(\xi) h_J(\xi) L_{\mathbf{rx}}(\xi) + \frac{2 h'_I(\xi) h'_J(\xi)}{l(\xi)} L_{\mathbf{rr}}(\xi) \quad ,$$
(F19)

$$\frac{\partial^2 \hat{L}}{\partial \mathbf{x}'_I \partial \mathbf{x}'_J} = \frac{l(\xi) d_I(\xi) d_J}{2}(\xi) L_{\mathbf{xx}}(\xi) + d_I(\xi) d'_J(\xi) L_{\mathbf{xr}}(\xi) \\ + d'_I(\xi) d_J(\xi) L_{\mathbf{rx}}(\xi) + \frac{2 d'_I(\xi) d'_J(\xi)}{l(\xi)} L_{\mathbf{rr}}(\xi) \quad ,$$
(F20)

$$\frac{\partial^2 \hat{L}}{\partial \mathbf{x}_I \partial \mathbf{x}'_J} = \frac{l(\xi) h_I(\xi) d_J(\xi)}{2} L_{\mathbf{xx}}(\xi) + h_I(\xi) d'_J(\xi) L_{\mathbf{xr}}(\xi) \\ + h'_I(\xi) d_J(\xi) L_{\mathbf{xr}}(\xi) + \frac{2 d'_I(\xi) d'_J(\xi)}{l(\xi)} L_{\mathbf{rr}}(\xi) \quad ,$$
(F21)

$$\frac{\partial^2 \hat{L}}{\partial \mathbf{x}'_I \partial \mathbf{x}_J} = \frac{l(\xi) d_I(\xi) h_J(\xi)}{2} L_{\mathbf{xx}}(\xi) + d_I(\xi) h'_J(\xi) L_{\mathbf{xr}}(\xi) \\ + d'_I(\xi) h_J(\xi) L_{\mathbf{rx}}(\xi) + \frac{2 d'_I(\xi) h'_J(\xi)}{l(\xi)} L_{\mathbf{rr}}(\xi) \quad .$$
(F22)

where $I$ is the row of the block, and $J$ is its column.

Since our degrees of freedom are $\mathbf{r}_I$ and $\mathbf{r}_J$ rather than $\mathbf{x}'_I$ and $\mathbf{x}'_J$, we apply (after integration of the traveltime gradient and Hessian blocks in $\xi$, over the finite element),

$$\frac{\partial \Delta t}{\partial \mathbf{r}_I} = \frac{l_I}{2} \frac{\partial \Delta t}{\partial \mathbf{x}'_I} \;, \quad \frac{\partial^2 \Delta t}{\partial \mathbf{x}_I \partial \mathbf{r}_J} = \frac{l_J}{2} \frac{\partial^2 \Delta t}{\partial \mathbf{x}_I \partial \mathbf{x}'_J} \;, \quad \frac{\partial^2 \Delta t}{\partial \mathbf{r}_I \partial \mathbf{x}_J} = \frac{l_I}{2} \frac{\partial^2 \Delta t}{\partial \mathbf{x}'_I \partial \mathbf{x}_J} \;, \quad \frac{\partial^2 \Delta t}{\partial \mathbf{r}_I \partial \mathbf{r}_J} = \frac{l_I l_J}{4} \frac{\partial^2 \Delta t}{\partial \mathbf{x}_I \partial \mathbf{x}'_J} \;, \quad (F23)$$

where $l_I/2$ and $l_J/2$ are the nodal values of the metric. Note that we do not compute the derivatives of the local traveltime Hessian components wrt $\xi$,

$$\frac{d}{d\xi}\frac{\partial^2 \hat{L}}{\partial \mathbf{x}_I \partial \mathbf{x}_J} \ , \ \frac{d}{d\xi}\frac{\partial^2 \hat{L}}{\partial \mathbf{x}'_I \partial \mathbf{x}'_J} \ , \ \frac{d}{d\xi}\frac{\partial^2 \hat{L}}{\partial \mathbf{x}_I \partial \mathbf{x}'_J} \ , \ \frac{d}{d\xi}\frac{\partial^2 \hat{L}}{\partial \mathbf{x}'_I \partial \mathbf{x}_J} \quad , \tag{F24}$$

which are needed at the ends of the interval for obtaining more accurate integration. Instead, we use the standard Simpson rule to integrate the local traveltime Hessian components. The reason is that these derivatives include the third derivatives of the velocity wrt locations and directions, and we do not want to assume that the medium properties are continuous up to the existence of the third derivatives ($C^2$ continuity). Note that in the Newton method, the accuracy is important mainly for the gradient. A small inaccuracy in the traveltime Hessian does not affect the accuracy of the result; it can only slightly affect the rate of convergence to the local stationary ray. However, on the last iteration, the traveltime Hessian for the stationary ray path should be computed accurately, as it is further used for dynamic ray tracing (DRT).

### APPENDIX G. FINITE-ELEMENT SOLVER FOR EULER-LAGRANGE EQUATION

In this appendix we show that the same algebraic equation set obtained by the direct search for the stationary traveltime (Appendix F) can be also obtained as a weak solution, with the Galerkin method, for the second-order, nonlinear Euler-Lagrange ODE derived in Part I. We start from equation 8 of Part I,

$$\frac{d}{ds}\left(\frac{\mathbf{r}}{v_{ray}} - \frac{\nabla_\mathbf{r} v_{ray}}{v_{ray}^2}\right) = -\frac{\nabla_\mathbf{x} v_{ray}}{v_{ray}^2} \quad . \tag{G1}$$

We multiply the Euler-Lagrange equation G1 by a weight (test) function $w(s)$ and integrate over the element length. The residual of the ODE is orthogonal to the test function,

$$\int_{s_{ini}}^{s_{fin}} \frac{d}{ds}\left(\frac{\mathbf{r}}{v_{ray}} - \frac{\nabla_\mathbf{r} v_{ray}}{v_{ray}^2}\right) w(s)\, ds = -\int_{s_{ini}}^{s_{fin}} \frac{\nabla_\mathbf{x} v_{ray}}{v_{ray}^2} w(s)\, ds \quad , \tag{G2}$$

where $s_{ini}$ and $s_{fin}$ are the values of the arclength at the endpoints of a finite element, $s_{ini} < s_{fin}$. Consider the left side of this equation and apply integration by parts,

$$\begin{aligned}
\int_{s_{ini}}^{s_{fin}} \frac{d}{ds}\left(\frac{\mathbf{r}}{v_{ray}} - \frac{\nabla_\mathbf{r} v_{ray}}{v_{ray}^2}\right) w\, ds &= \int_{s_{ini}}^{s_{fin}} w\, d\left(\frac{\mathbf{r}}{v_{ray}} - \frac{\nabla_\mathbf{r} v_{ray}}{v_{ray}^2}\right) \\
&= \left(\frac{\mathbf{r}}{v_{ray}} - \frac{\nabla_\mathbf{r} v_{ray}}{v_{ray}^2}\right) w \Bigg|_{s_{ini}}^{s_{fin}} - \int_{s_{ini}}^{s_{fin}} \left(\frac{\mathbf{r}}{v_{ray}} - \frac{\nabla_\mathbf{r} v_{ray}}{v_{ray}^2}\right) dw \\
&= \left(\frac{\mathbf{r}}{v_{ray}} - \frac{\nabla_\mathbf{r} v_{ray}}{v_{ray}^2}\right) w \Bigg|_{s_{ini}}^{s_{fin}} - \int_{s_{ini}}^{s_{fin}} \left(\frac{\mathbf{r}}{v_{ray}} - \frac{\nabla_\mathbf{r} v_{ray}}{v_{ray}^2}\right) \dot{w}\, ds
\end{aligned} \quad , \tag{G3}$$

where,

$$\dot{w} = \frac{dw}{ds} = \frac{dw}{d\xi}\frac{d\xi}{ds} = \frac{dw/d\xi}{ds/d\xi} = \frac{w'(\xi)}{s'(\xi)} \quad . \tag{G4}$$

Equation G2 becomes,

$$\left(\frac{\mathbf{r}}{v_{ray}} - \frac{\nabla_\mathbf{r} v_{ray}}{v_{ray}^2}\right) w \Bigg|_{s_{ini}}^{s_{fin}} - \int_{s_{ini}}^{s_{fin}} \left(\frac{\mathbf{r}}{v_{ray}} - \frac{\nabla_\mathbf{r} v_{ray}}{v_{ray}^2}\right) \dot{w}\, ds = -\int_{s_{ini}}^{s_{fin}} \frac{\nabla_\mathbf{x} v_{ray}}{v_{ray}^2} w\, ds \quad . \tag{G5}$$

According to equation A14 of Part I, the expression in the brackets is the slowness $\mathbf{p}$. We simplify the boundary term applying the abovementioned equation, but we keep the traveltime integrand (the Lagrangian) as in equation G5, because we derive the finite-element formulation in terms of the ray velocity and its derivatives (rather than in terms of the slowness).

Integration wrt the arclength can be replaced by integration wrt the internal parameter $\xi$,

$$\mathbf{p}\, w \Big|_{\xi=-1}^{\xi=+1} - \int_{\xi=-1}^{\xi=+1} \left( \frac{\mathbf{r}}{v_{\text{ray}}} - \frac{\nabla_{\mathbf{r}} v_{\text{ray}}}{v_{\text{ray}}^2} \right) w'\, d\xi = -\int_{\xi=-1}^{\xi=+1} \frac{\nabla_{\mathbf{x}} v_{\text{ray}}}{v_{\text{ray}}^2}\, w\, s'\, d\xi \quad , \tag{G6}$$

where $-1 \leq \xi \leq +1$ is the internal flow parameter within a single finite element, $s = s(\xi)$, $s(-1) = s_{\text{ini}}$, $s(+1) = s_{\text{fin}}$. The Galerkin method is widely used in the finite element analysis. The method suggests that the test (weight) functions $w(\xi)$ are the same as the interpolation (shape) functions. In our case these are the Hermite polynomials. There are $2n$ test functions per element, where $n = 2, 3$ is the number of nodes in a single element. These functions are $h_I(\xi)$ and $d_I(\xi)$, listed in equations A3 and A6 for two-node and three-node elements, respectively. At the joint nodes, $\xi = +1$ for the preceding element, and $\xi = -1$ for the subsequent element, and therefore, the boundary terms, being added on the assembly stage, cancel each other. Thus, the boundary condition (BC) terms appear at the ray path endpoints only (rather than at the finite-element endpoints). Furthermore, the interpolation functions $d_I(\xi)$ vanish at the ends of an element, $\xi = \pm 1$, and $h_I(\xi)$ accept value 1 at the corresponding end, and zero at the other end.

For each node $I$ of the finite element, we apply $w = h_I(\xi)$ and $w = d_I(\xi)$, and obtain two blocks (vectors) of length 3: the $h$-block related to spatial DoF and the $d$-block related to direction DoF. For the whole element, there are $2n$ blocks, or $6n$ scalar equations. The boundary term (the slowness vector with the corresponding sign on the left side of equation G6) appears with the minus sign for the source and with the plus sign for the receiver, in the corresponding $h$-blocks only. However, the $h$-block with the non-vanishing boundary term in equation G6 for the source and receiver elements can be removed and replaced by the specified endpoint locations,

$$\mathbf{x}_o = \mathbf{x}_S \quad , \quad \mathbf{x}_N = \mathbf{x}_R \quad , \tag{G7}$$

where $N+1$ is the total number of nodes in the finite-element scheme, enumerated from zero to $N$. Thus, equation G6 simplifies to,

$$\int_{\xi=-1}^{\xi=+1} \left( \frac{\mathbf{r}\, w'}{v_{\text{ray}}} - \frac{\nabla_{\mathbf{r}} v_{\text{ray}}\, w'}{v_{\text{ray}}^2} - \frac{\nabla_{\mathbf{x}} v_{\text{ray}}}{v_{\text{ray}}^2} w\, s' \right) d\xi = 0 \quad , \tag{G8}$$

for all internal DoF, and the BC equation G7 holds for the six external DoF. The boundary term has been dropped in equation G8, but it has no effect on the global traveltime variation due to cancellation at the element joints. The boundary terms should appear at the source and receiver only, but at these points they are removed and replaced by the endpoint location BC of equation G7.

Applying the spatial and directional shape functions, $w = h(\xi)$ and $w = d(\xi)$, and taking into account that,

$$\mathbf{r} = \dot{\mathbf{x}} = \frac{dx}{ds} = \frac{d\mathbf{x}/d\xi}{ds/d\xi} = \frac{\mathbf{x}'(\xi)}{s'(\xi)} = \frac{2}{l(\xi)}\mathbf{x}'(\xi) \tag{G9}$$

we obtain the final form of the weak formulation,

$$\int_{\xi=-1}^{\xi=+1}\left[\frac{h_I'(\xi)\mathbf{r}(\xi)}{v_{\text{ray}}} - \frac{h_I'(\xi)\nabla_{\mathbf{r}}v_{\text{ray}}}{v_{\text{ray}}^2} + \frac{l(\xi)h_I(\xi)\nabla_{\mathbf{x}}v_{\text{ray}}}{2v_{\text{ray}}^2}\right]d\xi = 0 ,$$

$$\int_{\xi=-1}^{\xi=+1}\left[\frac{d_I'(\xi)\mathbf{r}(\xi)}{v_{\text{ray}}} - \frac{d_I'(\xi)\nabla_{\mathbf{r}}v_{\text{ray}}}{v_{\text{ray}}^2} - \frac{l(\xi)d_I(\xi)\nabla_{\mathbf{x}}v_{\text{ray}}}{2v_{\text{ray}}^2}\right]d\xi = 0 . \tag{G10}$$

Taking into account the spatial and directional gradients of the arclength-related Lagrangian (equation set F2 of Part 1),

$$L_{\mathbf{x}} = -\frac{\nabla_{\mathbf{x}}v_{\text{ray}}}{v_{\text{ray}}^2} , \qquad L_{\mathbf{r}} = \frac{\mathbf{r}}{v_{\text{ray}}} - \frac{\nabla_{\mathbf{r}}v_{\text{ray}}}{v_{\text{ray}}^2} , \tag{G11}$$

we rearrange the weak formulation in equation set G10,

$$\int_{\xi=-1}^{\xi=+1}\left[\frac{l(\xi)h_I(\xi)}{2}L_{\mathbf{x}} + h_I'L_{\mathbf{r}}\right]d\xi = 0 , \quad \int_{\xi=-1}^{\xi=+1}\left[\frac{l(\xi)d_I(\xi)}{2}L_{\mathbf{x}} + d_I'L_{\mathbf{r}}\right]d\xi = 0 . \tag{G12}$$

We notice that the integrands on the left-hand side of equation set G12 coincide with the integrands $\partial L/\partial \mathbf{x}_I$ and $\partial L/\partial \mathbf{x}_I'$ in equations F8 and F9 for the spatial and directional sub-blocks of the local traveltime gradient. This means that our direct search for the stationary traveltime (presented in Appendix F) is equivalent to the weak solution of the second-order, nonlinear Euler-Lagrange ODE set, reduced to the first-order, nonlinear, local weighted residuals, with the use of the Galerkin method (equation set G12).

# APPENDIX H. CONSTRAINTS OF TRAVELTIME OPTIMIZATION

In addition to the traveltime, the target function includes two penalty terms: a term related to the location (distribution) of the nodes along the stationary path, and a term related to the normalization of the nodal ray velocity directions.

Node distribution penalty

The arclengths of any two successive intervals are inversely proportional to the average curvatures of the ray path within these intervals. The curvature is defined as,

$$\kappa = |\dot{\mathbf{r}}| = \left|\frac{d\mathbf{r}}{ds}\right| = \frac{|\mathbf{x}' \times \mathbf{x}''|}{|\mathbf{x}'|^3} \quad . \tag{H1}$$

We define the penalty terms for both two-node elements and three-node elements.

Two-node Hermite element

The mean (average) curvature of an interval reads,

$$\bar{\kappa} = \frac{1}{\Delta s}\int_{\xi=-1}^{\xi=+1}\kappa(\xi)ds = \frac{1}{\Delta s}\int_{-1}^{+1}\kappa(\xi)\frac{ds}{d\xi}d\xi = \frac{1}{2\Delta s}\int_{-1}^{+1}\kappa(\xi)l(\xi)d\xi = \frac{1}{\Delta s}\int_{-1}^{+1}\frac{|\mathbf{x}' \times \mathbf{x}''|}{|\mathbf{x}'|^2}d\xi \quad , \tag{H2}$$

where $\Delta s$ is the arclength of the element computed in the previous iteration. We assign the element lengths inversely proportional to the mean curvature. For $N$ finite elements constituting the path, this leads to,

$$\frac{1}{\bar{\kappa}_1} + \frac{1}{\bar{\kappa}_2} + \cdots + \frac{1}{\bar{\kappa}_n} = \frac{1}{\bar{\kappa}_M} \quad . \tag{H3}$$

The coefficients of proportionality, assigned to the element lengths, become,

$$\Lambda_i = \frac{\bar{\kappa}_M}{\bar{\kappa}_i} \ , \quad \sum_{i=1}^{N} \Lambda_i = 1 \quad . \tag{H4}$$

To avoid zero divide and zero length (zero traveltime) on the straight intervals, we modify equation H2,

$$\bar{\kappa} = \frac{1}{\Delta s} \int_{-1}^{+1} \frac{|\mathbf{x}' \times \mathbf{x}''|}{|\mathbf{x}'|^2} d\xi + d_c^{-1} \quad , \tag{H5}$$

where $d_c$ is a large characteristic distance (its reciprocal has to be small). This may, for example, be the distance between the trajectory endpoints, or that distance taken with factor 10. Thus, for each two successive elements labeled $i$ and $i + 1$, we need to minimize the difference between their scaled arclengths,

$$T = t + W_s + W_r \to \min \ , \quad W_s = \sum_{i=1}^{N-1} b_i = \frac{w_s}{2} \sum_{i=1}^{N-1} \left( \frac{\Delta s_i}{\Lambda_i} - \frac{\Delta s_{i+1}}{\Lambda_{i+1}} \right)^2 \quad , \tag{H6}$$

where $T$ is the target function, $t$ is the traveltime, $W_s$ is the node distribution penalty term, $w_s$ is its weight, $b_i$ is a penalty term related to a single joint, $W_r$ is the direction normalization penalty term (to be explained later), $N + 1$ is the total number of nodes, including two endpoints, $N$ is the number of two-node finite elements, and $N - 1$ is the number of joints (internal nodes). A constraint given by equation H6 contributes to the target function, its gradient and Hessian. We distinguish between the local and global gradients and Hessians of the constraint. Since the constraint is imposed on two successive elements, and each node has six DoF, the local constraint gradient is a vector of length $6(2n - 1)$, and the local Hessian is a square matrix of the same dimension, where $n$ is the number of nodes in a single finite element. This makes the dimensions 18 and 30 for two-node and three-node elements, respectively, and defines the band width of the global Hessian matrix. However, we will demonstrate later that for three-node elements, the constraint size can still be 18. The arclength, its gradient and Hessian are computed with the corresponding formulae for the traveltime and its derivatives, where we assign a unit to the ray velocity; thus, all velocity gradients and Hessians drop off from the formulae,

$$v_{\text{ray}} \equiv 1 \; , \quad \nabla_{\mathbf{x}} v_{\text{ray}} = \nabla_{\mathbf{r}} v_{\text{ray}} = \nabla_{\mathbf{x}} \nabla_{\mathbf{x}} v_{\text{ray}} = \nabla_{\mathbf{r}} \nabla_{\mathbf{r}} v_{\text{ray}} = \nabla_{\mathbf{x}} \nabla_{\mathbf{r}} v_{\text{ray}} = \nabla_{\mathbf{r}} \nabla_{\mathbf{x}} v_{\text{ray}} = 0 \quad . \quad \text{(H7)}$$

Let nodes $A$ and $B$ be related to the previous segment (two-node finite element) of the joint, while $B$ and $C$ – to the next. Node $B$ is the joint of the two successive segments. The arclength $\Delta s_i$ depends on location and directional coordinates of nodes $A$ and $B$, while the arclength $\Delta s_{i+1}$ – on the coordinates of nodes $B$ and $C$. The contribution of a single joint to the constraint penalty term follows from equation H6,

$$b_i = \frac{w_s}{2\Lambda_i^2} \Delta s_i^2 (\mathbf{d}_a, \mathbf{d}_b) - \frac{w_s}{\Lambda_i \Lambda_{i+1}} \Delta s_i (\mathbf{d}_a, \mathbf{d}_b) \Delta s_{i+1} (\mathbf{d}_b, \mathbf{d}_c) + \frac{w_s}{2\Lambda_{i+1}^2} \Delta s_{i+1}^2 (\mathbf{d}_b, \mathbf{d}_c) \quad , \quad \text{(H8)}$$

where $i$ is the number of the joint node connecting elements $i$ and $i+1$, and $\mathbf{d}_a, \mathbf{d}_b$ and $\mathbf{d}_c$ are degrees of freedom related to the corresponding nodes ($B$ is the joint node),

$$\mathbf{d}_a = [\mathbf{x}_a, \mathbf{r}_a] \; , \quad \mathbf{d}_b = [\mathbf{x}_b, \mathbf{r}_b] \; , \quad \mathbf{d}_c = [\mathbf{x}_c, \mathbf{r}_c] \quad . \quad \text{(H9)}$$

The gradient of this function consists of three blocks, each of length 6, and the Hessian consists of $3 \times 3$ blocks, where each block has dimensions $6 \times 6$.

Three-node Hermite element

For a three-node element, the joint between two successive intervals may be the central node of the element, or the joint between the two adjacent elements. The latter case is shown in Figure 12. Interval $i$ of arclength $\Delta s_i$ is the second half of element $k$, where $0 \leq \xi \leq +1$, and the adjacent interval $i+1$ of length $\Delta s_{i+1}$ is the first half of element $k+1$, where $-1 \leq \xi \leq 0$. In this case, we apply the joint node, the node to the left and the node to the right, in order to create an artificial finite element, shown in red. With this technique, the joint node is always the central node of the element, and the dimension of the constraint Hessian matrix is $3 \times 6 = 18$, rather than $5 \times 6 = 30$. This case, however, differs from the joint of a two-node element. In the case of a two-node element, the length $\Delta s_i$ of the "previous" interval $i$ depends on 12 DoF $\mathbf{d}_a, \mathbf{d}_b$, while the length $\Delta s_{i+1}$ of the "next" interval $i+1$ depends on 12 DoF $\mathbf{d}_b, \mathbf{d}_c$. In the case of a three-node element, the length $\Delta s_i$ of the "previous" interval $i$ depends on 18 DoF $\mathbf{d}_a, \mathbf{d}_b, \mathbf{d}_c$, while the length $\Delta s_{i+1}$ of the "next" interval $i+1$ depends the same 18 DoF, and equation H8 becomes,

$$b_i = \frac{w_s}{2\Lambda_i^2} \Delta s_i^2 (\mathbf{d}_a, \mathbf{d}_b, \mathbf{d}_c) + \frac{w_s}{2\Lambda_{i+1}^2} \Delta s_{i+1}^2 (\mathbf{d}_a, \mathbf{d}_b, \mathbf{d}_c) \\ - \frac{w_s}{\Lambda_i \Lambda_{i+1}} \Delta s_i (\mathbf{d}_a, \mathbf{d}_b, \mathbf{d}_c) \Delta s_{i+1} (\mathbf{d}_a, \mathbf{d}_b, \mathbf{d}_c) \quad . \tag{H10}$$

The local gradient and Hessian of the constraint, related to a single joint, become,

$$\frac{\partial b_i}{\partial \mathbf{u}} = +w_s \left( \frac{\Delta s_i}{\Lambda_i} - \frac{\Delta s_{i+1}}{\Lambda_{i+1}} \right) \left( \frac{1}{\Lambda_i} \frac{\partial \Delta s_i}{\partial \mathbf{u}} - \frac{1}{\Lambda_{i+1}} \frac{\partial \Delta s_{i+1}}{\partial \mathbf{u}} \right) \quad , \tag{H11}$$

$$\frac{\partial^2 b_i}{\partial \mathbf{u} \partial \mathbf{v}} = w_s \left( \frac{1}{\Lambda_i} \frac{\partial \Delta s_i}{\partial \mathbf{u}} - \frac{1}{\Lambda_{i+1}} \frac{\partial \Delta s_{i+1}}{\partial \mathbf{u}} \right) \otimes \left( \frac{1}{\Lambda_i} \frac{\partial \Delta s_i}{\partial \mathbf{v}} - \frac{1}{\Lambda_{i+1}} \frac{\partial \Delta s_{i+1}}{\partial \mathbf{v}} \right)$$
$$+ w_s \left( \frac{\Delta s_i}{\Lambda_i} - \frac{\Delta s_{i+1}}{\Lambda_{i+1}} \right) \left( \frac{1}{\Lambda_i} \frac{\partial^2 \Delta s_i}{\partial \mathbf{u} \partial \mathbf{v}} - \frac{1}{\Lambda_{i+1}} \frac{\partial^2 \Delta s_{i+1}}{\partial \mathbf{u} \partial \mathbf{v}} \right) ,$$
(H12)

where both vectors $\mathbf{u}$ and $\mathbf{v}$ should be replaced by $\mathbf{d}_a, \mathbf{d}_b, \mathbf{d}_c$. Equations H11 and H12 can be applied to two-node finite elements as well, where some terms of the gradient and Hessian vanish, because in this case $\partial \Delta_k / \partial \mathbf{d}_c = \partial \Delta_{k+1} / \partial \mathbf{d}_a = 0$.

Ray direction normalization penalty

Another penalty term is related to the unit length of the nodal ray velocity directions,

$$W_r = \sum_{i=1}^{N} w_{r,i} = \frac{w_r}{2} \sum_{i=0}^{N} (\mathbf{r}_i \cdot \mathbf{r}_i - 1)^2 , \qquad (H13)$$

where $W_r$ is the penalty term accounting for all the nodes, $w_{r,i}$ is the contribution to of a single node to this term, and $N + 1$ is the total number of nodes, including the two end nodes and enumerated from zero to $N$. The local gradient of the constraint has length 3, and the local Hessian is a matrix of dimensions $3 \times 3$,

$$\frac{\partial w_{r,i}}{\partial \mathbf{r}_i} = 2 w_r (\mathbf{r}_i \cdot \mathbf{r}_i - 1) \mathbf{r} , \quad \frac{\partial^2 w_{r,i}}{\partial \mathbf{r}_i^2} = 4 w_r \mathbf{r}_i \otimes \mathbf{r}_i + 2 w_r (\mathbf{r}_i \cdot \mathbf{r}_i - 1) \mathbf{I} . \qquad (H14)$$

An alternative approach for keeping the ray velocity direction normalized is to apply the "hard" constraint with Lagrange multipliers (instead of the soft constraint with the penalty term added to the target function). This leads to,

$$W_r = \sum_{i=1}^{N} w_{r,i} = \sum_{i=1}^{N} \lambda_i \left( \mathbf{r}_i \cdot \mathbf{r}_i - 1 \right) \qquad . \qquad \text{(H15)}$$

This is a more accurate approach that enforces the strict normalization, but it requires seven DoF per node: $\mathbf{x}_i, \mathbf{r}_i$ and $\lambda_i$, and increases both the size of the global Hessian matrix and the bandwidth. In the numerical examples that we demonstrate, the soft constraints work fine.

# APPENDIX I. ASSEMBLY OF THE TRAVELTIME HESSIAN AND GRADIENT AND SETTING THE BOUNDARY CONDITIONS

In this appendix we explain the assembly procedure and implementation of the boundary conditions with an example. Consider a ray path consisting of three three-node elements as shown in Figure 13. The nodes have sequential global numeration. The scheme includes three elements, seven nodes and two joints. Nodes 0 and 6 are end nodes, and their locations are known.

Assembly.

The assembly of the traveltime Hessian matrix and gradient vector is shown in Figure 14. Each cell in the gradient is a vector of length 6. Each cell in the Hessian is a matrix block of dimensions $6 \times 6$. Each element yields a local vector of length 18 and a local matrix of $18 \times 18$. Yellow, green and light blue show the contributions of the three elements, and the overlaps are shown in red. Contributions of neighboring elements are added in the overlap cells. Symbol $\mathbf{d}_i$ in Figure 14 means both location and direction DoF of node $i$, $\mathbf{d}_i = [\mathbf{x}_i, \mathbf{r}_i]$.

The penalty terms of the constraints included within the target function result in their own gradients and Hessians. The assembly of the local gradients and Hessians of the node distribution constraint is shown in Figure 15. A penalty term related to each internal node regulates the arclengths of two adjacent segments of the path. In case of two-node elements, these segments belong to the neighbor elements connected at the node. In case of three-node elements, the segments may belong either to the same element, or to the second "half" of the previous element and the first "half" of the next element that meet at the joint node. (Two segments of three-node element have generally different arclengths.) In either case, the constraint term describes two segments related to three nodes, and therefore, it yields a local gradient vector of length 18 and a local Hessian matrix of dimension $18 \times 18$ (regardless of the order of the elements: two-node or three-node) which contribute to the global gradient and Hessian of the constraint. Different colors in Figure 15 show the contributions of the internal nodes; again, the overlap is in red.

Contributions of the normalization constraint are shown in Figure 16. The light blue cells are vectors of length 3 and matrix blocks $3 \times 3$.

Next, we add the global gradients and Hessians of the traveltime and two constraints to obtain the global gradient and Hessian of the target function $T$, respectively.

Discussion about assembly of models with discontinuous ray velocity

In this study we assume a smooth velocity (elastic tensor) field, normally represented on a 3D grid. The interpolation method along the ray trajectory guarantees $C^1$ continuity between finite elements, which means continuity of the function (ray point location) and its derivative (ray point direction). Note that "blocky" models with sharp discontinuities across the interfaces result in jumps of the ray velocity at these points (both direction and magnitude) and require a different dedicated strategy. Although this topic is beyond the scope of this study, the proposed method can be extended to include discontinuities of the ray direction at some specific nodes as well. This is one of the advantages in using the Hermite-type finite elements. We briefly discuss this option below.

In order to model discontinuous ray velocity directions, we first identify (or define) the "discontinues" nodes and we add to each of them a set of three extra global direction DoF (Cartesian components). Note that for the ray paths with discontinuous ray velocity directions, the computation of the local traveltime gradient vectors and Hessian matrices does not change; only the assembly procedure is different. At the interface joints with the extra direction DoF, the location DoF overlap, and the values of the local gradients and Hessians are added in their common cell. Due to the three extra DoF, the direction and mixed cells of the adjacent elements are different and do not overlap.

Assume for simplicity's sake that there are only two finite elements, and two nodes per element (12 DoF per element). Since the location DoF are common for the two elements and the direction DoF are not, the total number of DoF is 21. The common (joint) node of these two elements belongs, for example, to the interface. This special node has one set (three Cartesian components) of location DoF and two sets of direction DoF.

Let $A, B$ and $C, D$ be the nodes of elements 1 and 2, respectively. Nodes $B$ and $C$ share their common location but have different ray directions. Assume the following order of the global

DoF: a) location of $A$, b) ray direction at $A$, c) common location of $B$ and $C$, $\mathbf{x}_B = \mathbf{x}_C$, d) direction at $B$, e) direction at $C$, f) location at $D$, and g) direction at $D$.

The assembly schemes for the global traveltime gradient and Hessian are shown in Tables 6 and 7, where the cells are vectors and square matrices, respectively, of dimension 3. The yellow background means summation of the content in the overlapped cells.

Models with discontinuous velocities, represented by surface interfaces, require additional constraints enforcing that the nodes with the double sets of direction DoF are located at these interfaces.

Boundary conditions

The locations of the end nodes are specified. The Newton method works with increments of the nodal locations and directions, so the increments of the end node locations are zero. The implementation of the boundary conditions is shown in Figure 17, where the white background means that these blocks are unchanged. The yellow background means cleaned blocks (all zeros). These are the first and the last-but-one sub-blocks of length 3 of the gradient vector and the corresponding row and column blocks of the Hessian matrix. The two green cells labeled $\mathbf{I}$ mean $3 \times 3$ identity matrices.

## APPENDIX J. NEWTON OPTIMIZATION

In each iteration of the Newton method, we refine the ray trajectory in order to decrease the target function $T(\mathbf{d})$. In the case of a minimum traveltime, the target function includes the traveltime and the penalty terms. In a more general stationary time case, which may be also a saddle point, the target function to be minimized includes the traveltime gradient squared and the penalty terms. At the minimum of the target function, the traveltime gradient should be zero or very small.

For the constrained minimum traveltime search, we apply the Newton method. For a constrained saddle-point search or in a general case where the type of the stationary path is unknown ahead of the iterative process, the gradient of the target function to be minimized already includes the traveltime Hessian, and we apply the gradient-based methods (like conjugate-gradient and anti-gradient descent).

When the Newton method is used to refine (update) the ray trajectory, we solve the linearized equation set that delivers the increments for the nodal locations and directions,

$$\nabla_{\mathbf{d}}\nabla_{\mathbf{d}}T \cdot \Delta \mathbf{d} = -\nabla_{\mathbf{d}}T \quad \text{or} \quad T_{\mathbf{dd}} \cdot \Delta \mathbf{d} = -T_{\mathbf{d}} \quad , \tag{J1}$$

where $\nabla_{\mathbf{d}}T$ and $\nabla_{\mathbf{d}}\nabla_{\mathbf{d}}T$ are the global gradient and Hessian of the target function, and vector $\mathbf{d}$ represents the nodal DoF (locations $\mathbf{x}_i$ and directions $\mathbf{r}_i$),

$$\mathbf{d} = [\mathbf{d}_o, \mathbf{d}_1, \ldots \mathbf{d}_N] \quad , \quad \mathbf{d}_i = [\mathbf{x}_i, \mathbf{r}_i] \quad . \tag{J2}$$

The refined DoF read,

$$\mathbf{d}_i^{(k+1)} = \mathbf{d}_i^{(k)} + \Delta \mathbf{d}_i \quad , \quad \Delta \mathbf{d}_i = [\Delta \mathbf{x}_i, \Delta \mathbf{r}_i] \quad , \tag{J3}$$

where $k$ is the iteration number. For each node, a short increment $\Delta \mathbf{r}_i$ is almost normal to $\mathbf{r}_i$ because this vector can only rotate, it cannot change its unit length.

When computing a partial derivative of the traveltime and other items of the target function $T$ wrt a directional component of a node, we assume that the two other directional components are fixed, for example,

$$\frac{\partial T}{\partial r_{i,1}} = \lim_{\Delta r_{i,1} \to 0} \frac{T(r_{i,1} + \Delta r_{i,1}, r_{i,2}, r_{i,3}) - T(r_{i,1}, r_{i,2}, r_{i,3})}{\Delta r_{i,1}} \quad , \tag{J4}$$

where $i$ is the node index, and indices 1, 2, 3 mean Cartesian components. However, we cannot change one directional component without simultaneously changing the other two components, as this ruins the normalization,

$$|\mathbf{r}| = 1 \quad , \quad |\mathbf{r} + \Delta \mathbf{r}| \neq 1 \quad . \tag{J5}$$

Breaking the normalization for directional derivatives of the target function leads to inconsistent results, unless we convert the non-normalized directional derivative (a vector consisting of three regular partial derivatives) into the normalized one (Ravve and Koren, 2019),

$$\left[ \frac{\partial T}{\partial \mathbf{r}_i} \right]^{\text{norm}} = (\mathbf{I} - \mathbf{r}_i \otimes \mathbf{r}_i) \left[ \frac{\partial T}{\partial \mathbf{r}_i} \right]^{\text{non-norm}} \quad , \tag{J6}$$

where $i$ is the node index. The optimization procedure enforces descent of the target function. The Newton method does not always converge, and therefore before updating the trial path we check that this update indeed reduces the target function,

$$T(\mathbf{d} + \Delta \mathbf{d}) < T(\mathbf{d}) \quad . \tag{J7}$$

In cases where the criterion of equation J7 is violated, we apply the counter-gradient descent,

$$\Delta \mathbf{d} = -s_{\text{ag}} T_{\mathbf{d}} = -s_{\text{ag}} \nabla_{\mathbf{d}} T \quad , \tag{J8}$$

where the optimum step is given by,

$$s_{\text{ag}} = \frac{T_{\mathbf{dd}}}{T_{\mathbf{d}} \cdot T_{\mathbf{dd}} \cdot T_{\mathbf{d}}} = \frac{\nabla_{\mathbf{d}} T \cdot \nabla_{\mathbf{d}} T}{\nabla_{\mathbf{d}} T \cdot \nabla_{\mathbf{d}} \nabla_{\mathbf{d}} T \cdot \nabla_{\mathbf{d}} T} \quad , \tag{J9}$$

where subscript 'ag' means anti-gradient. The numerator is always positive, while the denominator is positive if the global Hessian $\nabla_{\mathbf{d}} \nabla_{\mathbf{d}} T$ is positive definite. The initial guess of the trial trajectory may still be far from the target function minimum, so we replace step $s_{\text{ag}}$ by its absolute value. We check condition J7 again, and if it does not hold, we decrease the step by factor 2. The step reduction can be repeated if necessary, until J7 holds. For some small steps in the counter-gradient direction the function always decreases, but this step may prove to be smaller than the optimum value provided by equation J9.

The local and global Hessians of the target function are symmetric matrices, and the global Hessian is also a narrow-band matrix. In the proximity of the target function minimum, the global Hessian is positive-definite, but far from the minimum this is not a must. The linearized equation set J1 is solved with the Cholesky decomposition where the symmetric matrix is factorized into the lower triangular, the diagonal and the upper triangular matrices. The two triangular matrices are mirrors of each other, and the diagonal matrix may include integers $+1$ and $-1$. In the proximity of the minimum, the Hessian of the target function becomes positive definite, and the diagonal matrix consists of $+1$ only, i.e., becomes the identity matrix, making it possible to check the minimum criterion without computing the eigenvalues of the global Hessian.

Numerical procedures for locating constrained saddle points have been studied by Zhang and Du (2012a, 2012b), Ren and Vanden-Eijnden (2013), Gao et al. (2015), Albareda et al. (2018), Li et al. (2019), Li and Zhou (2019).

# LIST OF TABLES

Table 1. Local traveltime gradient, includes $n$ blocks.

| |
|---|
| $A$ |
| $B$ |
| $C$ |

Table 2. Local traveltime gradient block.

| | |
|---|---|
| $\dfrac{d\Delta t}{dx_{I,1}}$ | Derivatives wrt location |
| $\dfrac{d\Delta t}{dx_{I,2}}$ | |
| $\dfrac{d\Delta t}{dx_{I,3}}$ | |
| $\dfrac{d\Delta t}{dr_{I,1}}$ | Derivatives wrt direction |
| $\dfrac{d\Delta t}{dr_{I,2}}$ | |
| $\dfrac{d\Delta t}{dr_{I,3}}$ | |

Table 3. Local traveltime Hessian matrix, includes $n \times n$ blocks.

| | | |
|---|---|---|
| $AA$ | $AB$ | $AC$ |
| $BA$ | $BB$ | $BC$ |
| $CA$ | $CB$ | $CC$ |

Table 4. Local traveltime Hessian block, each cell is a 3 × 3 sub-block (tensor).

| Second derivatives of traveltime wrt nodal locations | Second derivatives of traveltime wrt nodal locations (first) and directions (second) |
|---|---|
| Second derivatives of traveltime wrt nodal directions (first) and locations (second) | Second derivatives of traveltime wrt nodal directions |

Table 5. Local traveltime Hessian block, each cell is a scalar.

| $\dfrac{\partial \Delta t}{\partial x_{I,1} \partial x_{J,1}}$ | $\dfrac{\partial \Delta t}{\partial x_{I,1} \partial x_{J,2}}$ | $\dfrac{\partial \Delta t}{\partial x_{I,1} \partial x_{J,3}}$ | $\dfrac{\partial \Delta t}{\partial x_{I,1} \partial r_{J,1}}$ | $\dfrac{\partial \Delta t}{\partial x_{I,1} \partial r_{J,2}}$ | $\dfrac{\partial \Delta t}{\partial x_{I,1} \partial r_{J,3}}$ |
|---|---|---|---|---|---|
| $\dfrac{\partial \Delta t}{\partial x_{I,2} \partial x_{J,1}}$ | $\dfrac{\partial \Delta t}{\partial x_{I,2} \partial x_{J,2}}$ | $\dfrac{\partial \Delta t}{\partial x_{I,2} \partial x_{J,3}}$ | $\dfrac{\partial \Delta t}{\partial x_{I,2} \partial r_{J,1}}$ | $\dfrac{\partial \Delta t}{\partial x_{I,1} \partial r_{J,2}}$ | $\dfrac{\partial \Delta t}{\partial x_{I,1} \partial r_{J,3}}$ |
| $\dfrac{\partial \Delta t}{\partial x_{I,3} \partial x_{J,1}}$ | $\dfrac{\partial \Delta t}{\partial x_{I,1} \partial x_{J,2}}$ | $\dfrac{\partial \Delta t}{\partial x_{I,3} \partial x_{J,3}}$ | $\dfrac{\partial \Delta t}{\partial x_{I,3} \partial r_{J,1}}$ | $\dfrac{\partial \Delta t}{\partial x_{I,1} \partial r_{J,2}}$ | $\dfrac{\partial \Delta t}{\partial x_{I,1} \partial r_{J,3}}$ |
| $\dfrac{\partial \Delta t}{\partial r_{I,1} \partial x_{J,1}}$ | $\dfrac{\partial \Delta t}{\partial r_{I,1} \partial x_{J,2}}$ | $\dfrac{\partial \Delta t}{\partial r_{I,1} \partial x_{J,3}}$ | $\dfrac{\partial \Delta t}{\partial r_{I,1} \partial r_{J,1}}$ | $\dfrac{\partial \Delta t}{\partial r_{I,1} \partial r_{J,2}}$ | $\dfrac{\partial \Delta t}{\partial r_{I,1} \partial r_{J,3}}$ |
| $\dfrac{\partial \Delta t}{\partial r_{I,2} \partial x_{J,1}}$ | $\dfrac{\partial \Delta t}{\partial r_{I,2} \partial x_{J,2}}$ | $\dfrac{\partial \Delta t}{\partial r_{I,2} \partial x_{J,3}}$ | $\dfrac{\partial \Delta t}{\partial r_{I,2} \partial r_{J,1}}$ | $\dfrac{\partial \Delta t}{\partial r_{I,2} \partial r_{J,2}}$ | $\dfrac{\partial \Delta t}{\partial r_{I,2} \partial r_{J,3}}$ |
| $\dfrac{\partial \Delta t}{\partial r_{I,3} \partial x_{J,1}}$ | $\dfrac{\partial \Delta t}{\partial r_{I,3} \partial x_{J,2}}$ | $\dfrac{\partial \Delta t}{\partial r_{I,3} \partial x_{J,3}}$ | $\dfrac{\partial \Delta t}{\partial r_{I,3} \partial r_{J,1}}$ | $\dfrac{\partial \Delta t}{\partial r_{I,3} \partial r_{J,2}}$ | $\dfrac{\partial \Delta t}{\partial r_{I,3} \partial r_{J,3}}$ |

Table 6. Assembly of global traveltime gradient for a model with discontinuous ray direction.

| DoF | Gradient |
|---|---|
| $\mathbf{x}_A$ | $\partial t / \partial \mathbf{x}_A$ |
| $\mathbf{r}_A$ | $\partial t / \partial \mathbf{r}_A$ |
| $\mathbf{x}_B = \mathbf{x}_C$ | $\partial t / \partial \mathbf{x}_B + \partial t / \partial \mathbf{x}_C$ |
| $\mathbf{r}_B$ | $\partial t / \partial \mathbf{r}_B$ |
| $\mathbf{r}_C$ | $\partial t / \partial \mathbf{r}_C$ |
| $\mathbf{x}_D$ | $\partial t / \partial \mathbf{x}_D$ |
| $\mathbf{r}_D$ | $\partial t / \partial \mathbf{r}_D$ |

Table 7. Assembly of global traveltime Hessian for a model with discontinuous ray direction.

| DoF | $\mathbf{x}_A$ | $\mathbf{r}_A$ | $\mathbf{x}_B = \mathbf{x}_C$ | $\mathbf{r}_B$ | $\mathbf{r}_C$ | $\mathbf{x}_D$ | $\mathbf{r}_D$ |
|---|---|---|---|---|---|---|---|
| $\mathbf{x}_A$ | $\dfrac{\partial^2 t}{\partial \mathbf{x}_A^2}$ | $\dfrac{\partial^2 t}{\partial \mathbf{x}_A \partial \mathbf{r}_A}$ | $\dfrac{\partial^2 t}{\partial \mathbf{x}_A \partial \mathbf{x}_B}$ | $\dfrac{\partial^2 t}{\partial \mathbf{x}_A \partial \mathbf{r}_B}$ | | | |
| $\mathbf{r}_A$ | $\dfrac{\partial^2 t}{\partial \mathbf{r}_A \partial \mathbf{x}_A}$ | $\dfrac{\partial^2 t}{\partial \mathbf{r}_A^2}$ | $\dfrac{\partial^2 t}{\partial \mathbf{r}_A \partial \mathbf{x}_B}$ | $\dfrac{\partial^2 t}{\partial \mathbf{r}_A \partial \mathbf{r}_B}$ | | | |
| $\mathbf{x}_B = \mathbf{x}_C$ | $\dfrac{\partial^2 t}{\partial \mathbf{x}_B \partial \mathbf{x}_A}$ | $\dfrac{\partial^2 t}{\partial \mathbf{x}_B \partial \mathbf{r}_A}$ | $\dfrac{\partial^2 t}{\partial \mathbf{x}_B^2} + \dfrac{\partial^2 t}{\partial \mathbf{x}_C^2}$ | $\dfrac{\partial^2 t}{\partial \mathbf{x}_B \partial \mathbf{r}_B}$ | $\dfrac{\partial^2 t}{\partial \mathbf{x}_C \partial \mathbf{r}_C}$ | $\dfrac{\partial^2 t}{\partial \mathbf{x}_C \partial \mathbf{x}_D}$ | $\dfrac{\partial^2 t}{\partial \mathbf{x}_C \partial \mathbf{r}_D}$ |
| $\mathbf{r}_B$ | $\dfrac{\partial^2 t}{\partial \mathbf{r}_B \partial \mathbf{x}_A}$ | $\dfrac{\partial^2 t}{\partial \mathbf{r}_B \partial \mathbf{x}_A}$ | $\dfrac{\partial^2 t}{\partial \mathbf{r}_B \partial \mathbf{x}_B}$ | $\dfrac{\partial^2 t}{\partial \mathbf{r}_B^2}$ | | | |
| $\mathbf{r}_C$ | | | $\dfrac{\partial^2 t}{\partial \mathbf{r}_C \partial \mathbf{x}_C}$ | | $\dfrac{\partial^2 t}{\partial \mathbf{r}_C^2}$ | $\dfrac{\partial^2 t}{\partial \mathbf{r}_C \partial \mathbf{x}_D}$ | $\dfrac{\partial^2 t}{\partial \mathbf{r}_C \partial \mathbf{r}_D}$ |
| $\mathbf{x}_D$ | | | $\dfrac{\partial^2 t}{\partial \mathbf{x}_D \partial \mathbf{x}_C}$ | | $\dfrac{\partial^2 t}{\partial \mathbf{x}_D \partial \mathbf{r}_C}$ | $\dfrac{\partial^2 t}{\partial \mathbf{x}_D^2}$ | $\dfrac{\partial^2 t}{\partial \mathbf{x}_D \partial \mathbf{r}_D}$ |
| $\mathbf{r}_D$ | | | $\dfrac{\partial^2 t}{\partial \mathbf{r}_D \partial \mathbf{x}_C}$ | | $\dfrac{\partial^2 t}{\partial \mathbf{r}_D \partial \mathbf{r}_C}$ | $\dfrac{\partial^2 t}{\partial \mathbf{r}_D \partial \mathbf{x}_D}$ | $\dfrac{\partial^2 t}{\partial \mathbf{r}_D^2}$ |

# LIST OF FIGURES

Figure 13. Ray path scheme.

Figure 14. Contribution of traveltime derivatives to the global gradient and Hessian.

Figure 15. Contribution of node distribution penalty to the global gradient and Hessian.

Figure 16. Contribution of normalization penalty to the global gradient and Hessian.

Figure 17. Implementation of boundary conditions.

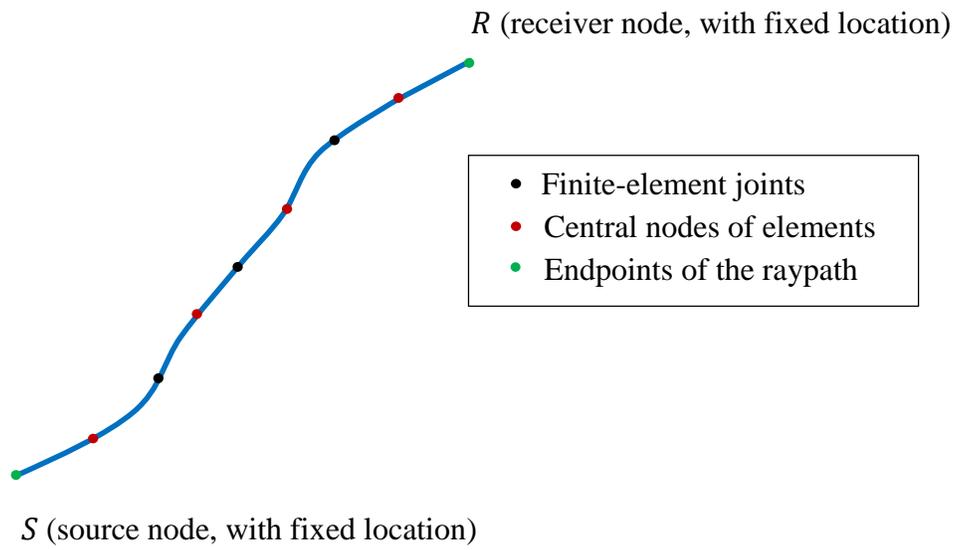

Figure 1. Discretization scheme of a ray path with three-node finite elements.

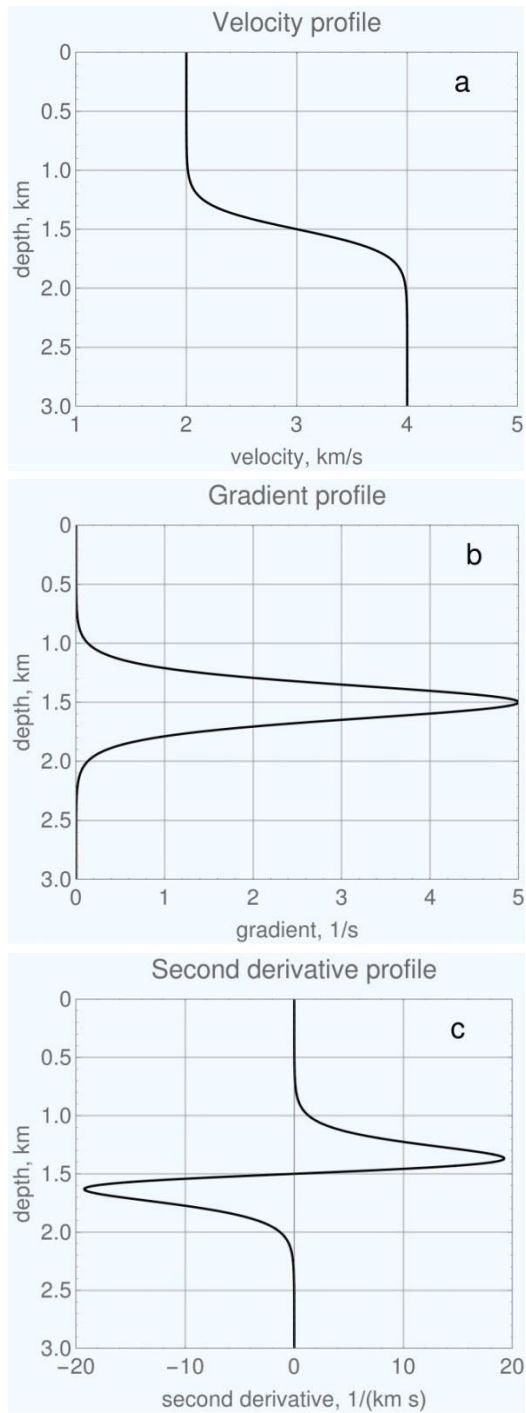

Figure 2. High-velocity half-space under a constant velocity layer: a) vertical velocity profile, b) vertical velocity gradient, c) second vertical derivative of the velocity.

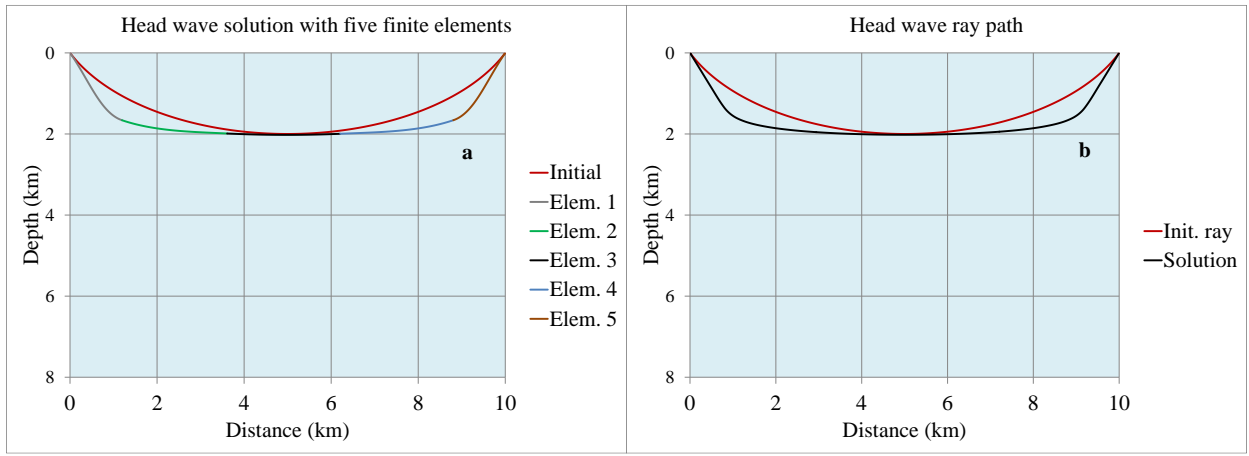

Figure 3. Head wave ray path in the medium with a high-velocity half-space: a) Eigenray with five finite elements, b) Eigenray with twenty finite elements.

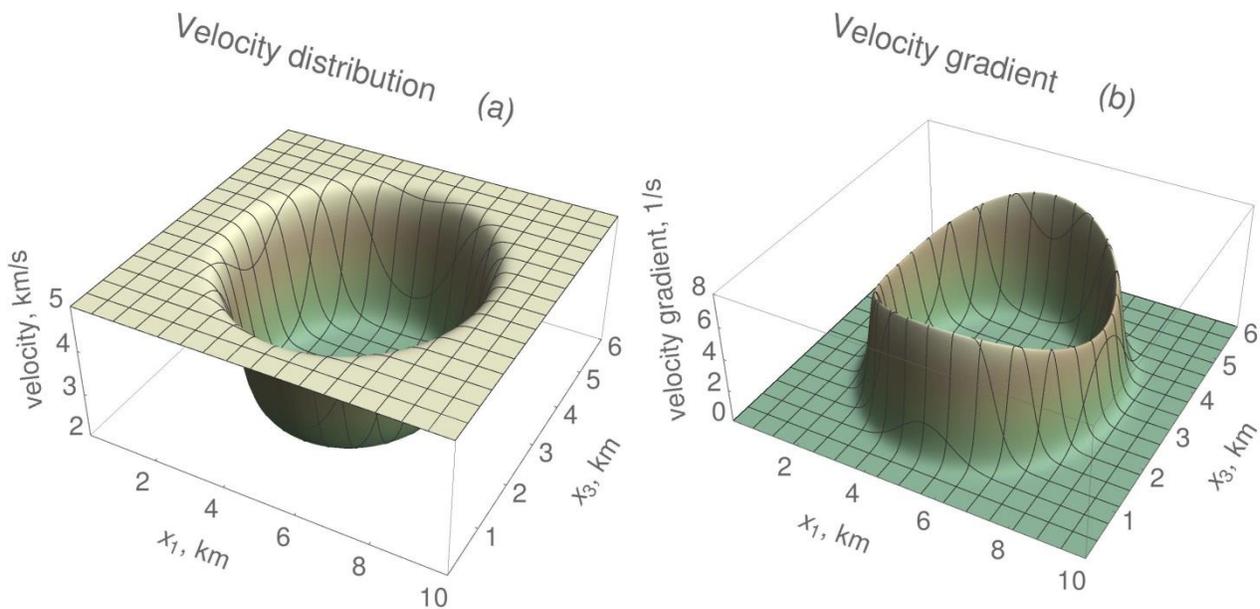

Figure 4. Medium with a low-velocity elliptic anomaly: a) velocity distribution, b) absolute value of the velocity gradient.

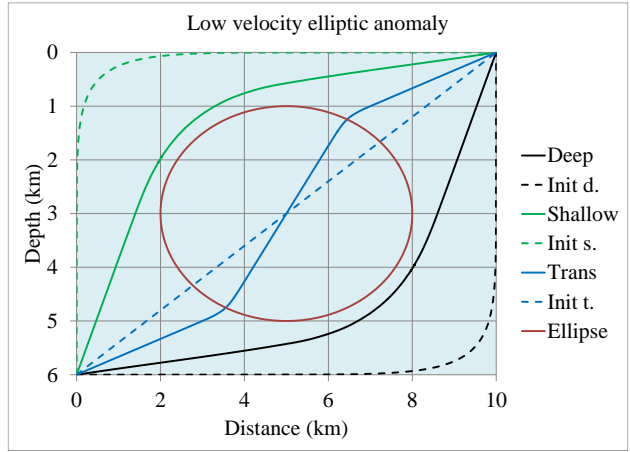

Figure 5. One-way path in a medium with low-velocity elliptic anomaly: Multi-arrival, corresponding to the shallow, deep and transmission solutions.

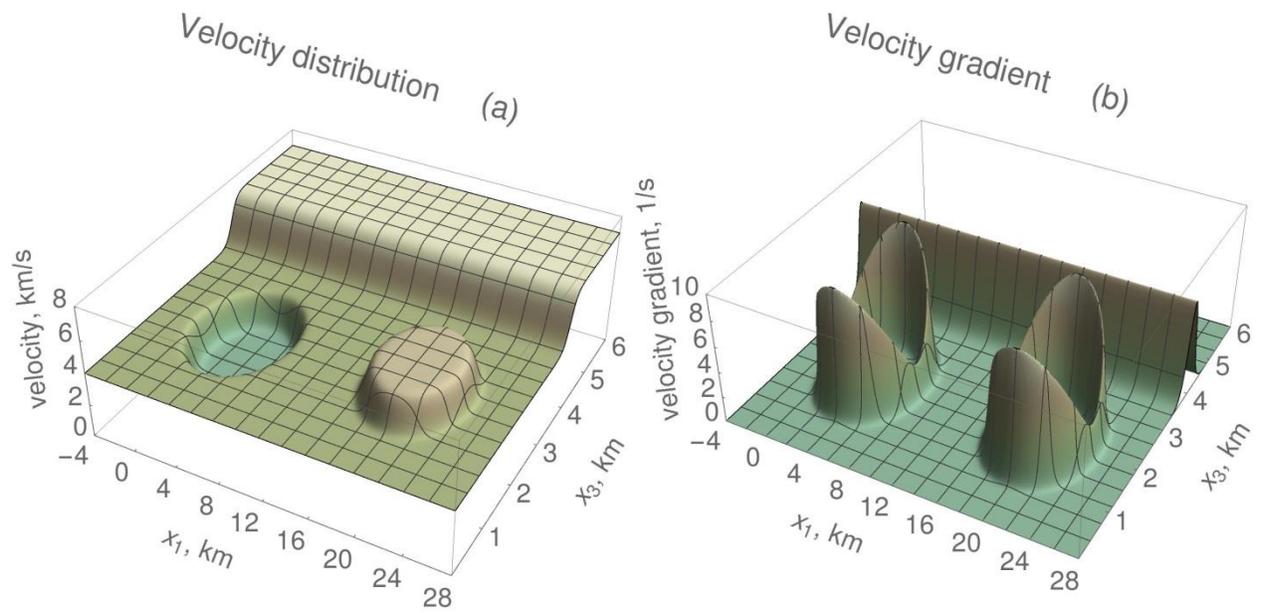

Figure 6. Medium with two elliptic anomalies and a high-velocity half-space: a) velocity distribution, b) absolute value of the velocity gradient.

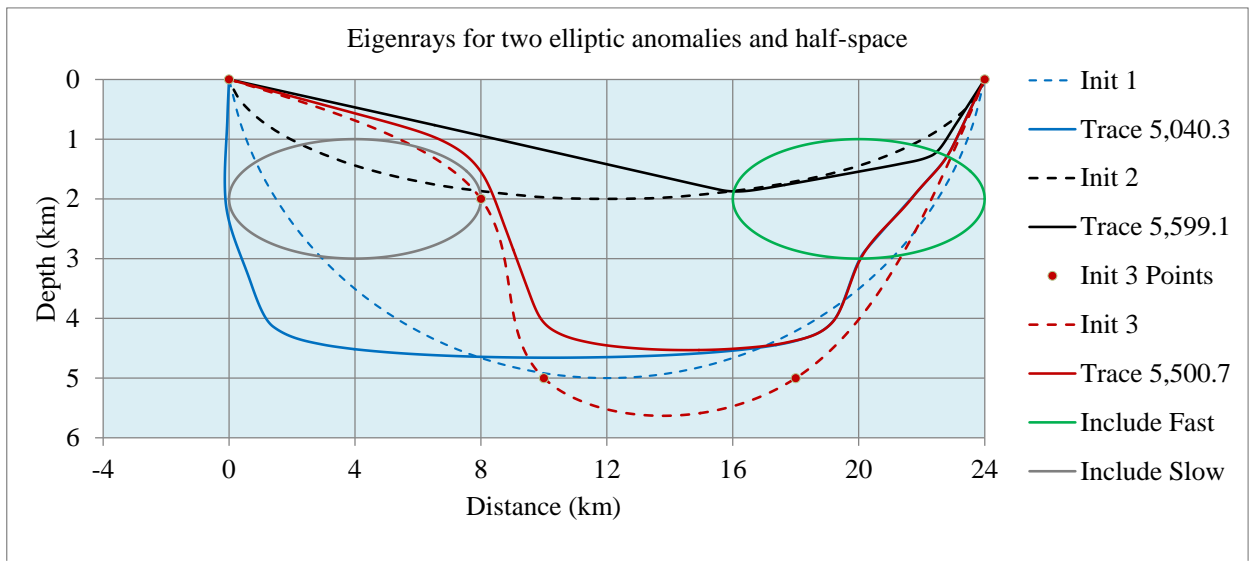

Figure 7. Eigenrays in a medium with two elliptic anomalies and a half-space.

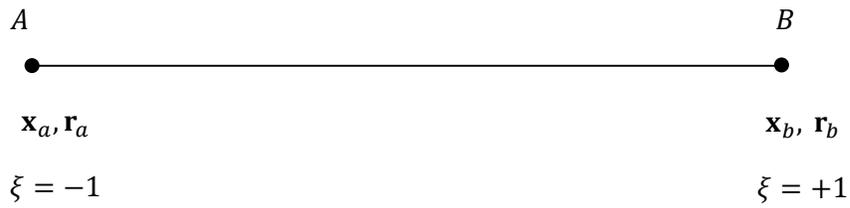

Figure 8. Scheme of two-node Hermite element.

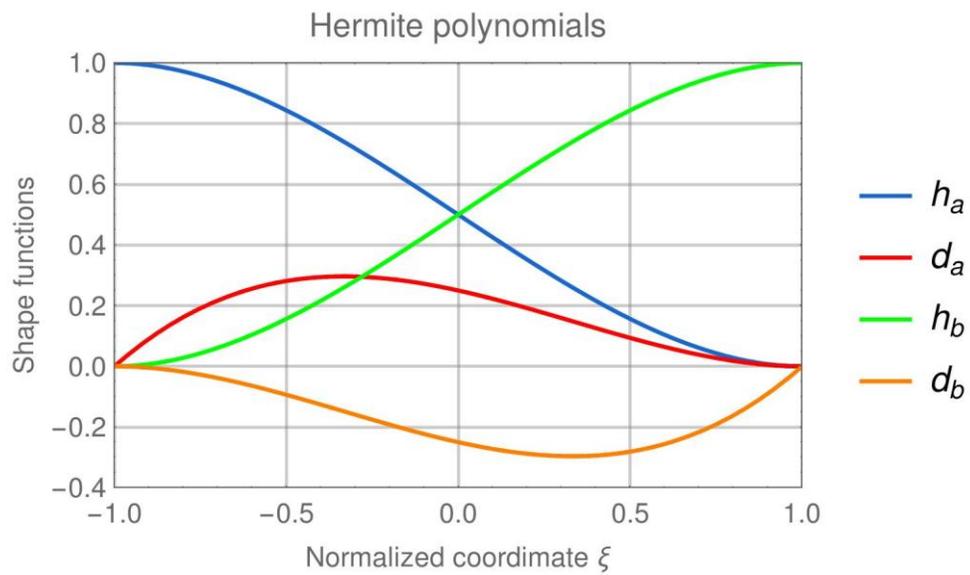

Figure 9. Shape functions for two-node Hermite interpolation.

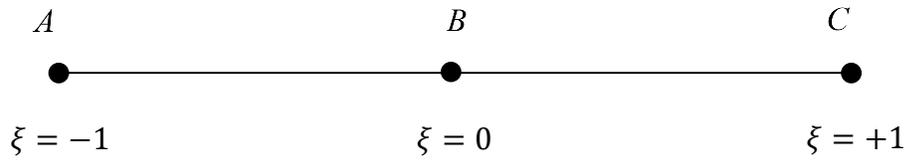

Figure 10. Scheme of three-node Hermite interpolation.

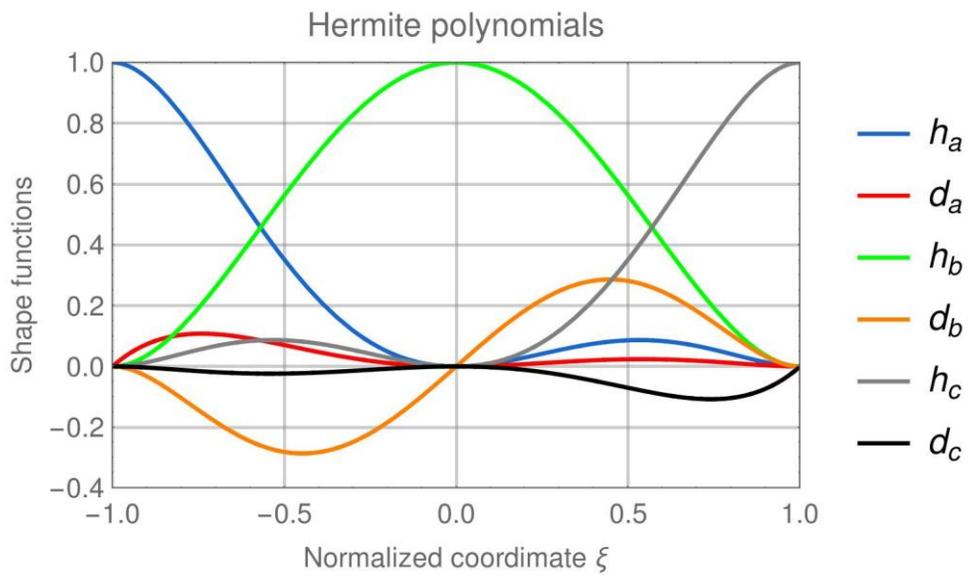

Figure 11. Shape functions for three-node Hermite interpolation.

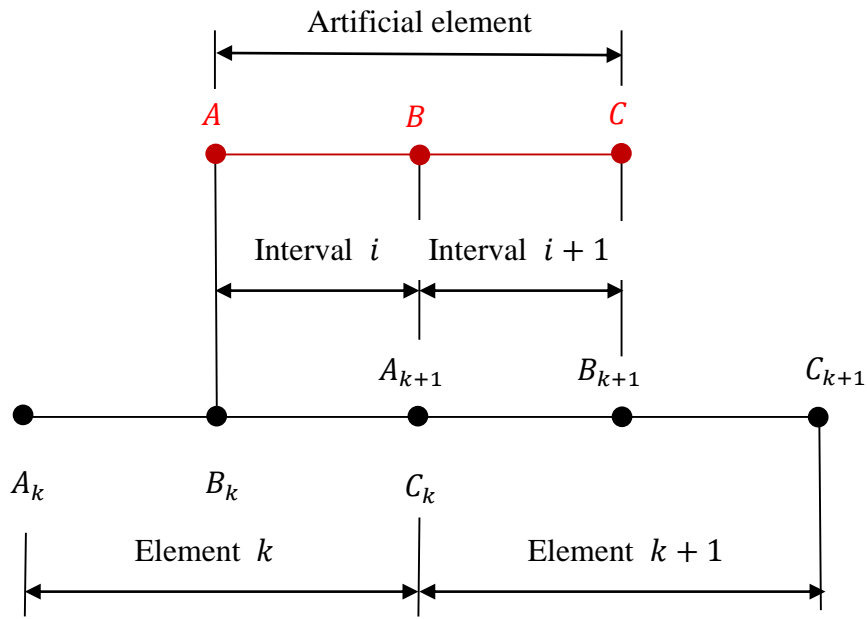

Figure 12. Joint between three-node elements.

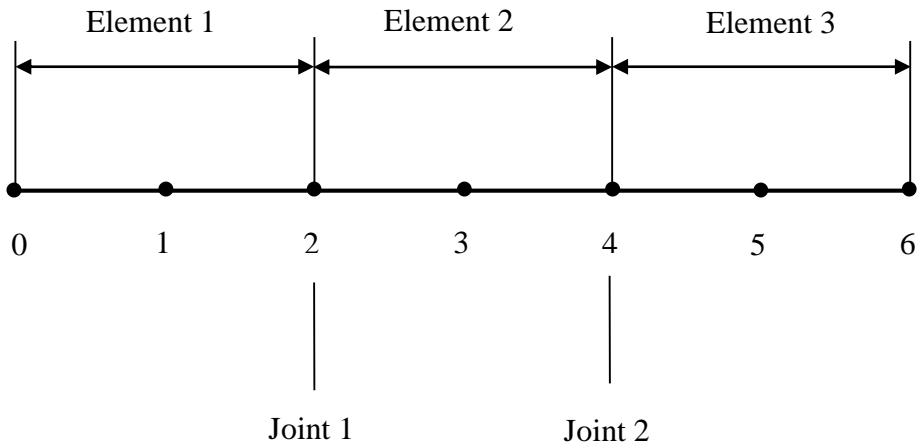

Figure 13. Ray path scheme.

Figure 14. Contribution of traveltime derivatives to the global gradient and Hessian.

Figure 15. Contribution of node distribution penalty to the global gradient and Hessian.

Figure 16. Contribution of normalization penalty to the global gradient and Hessian.

Figure 17. Implementation of boundary conditions.